\documentclass[journal,twoside]{IEEEtran}
\usepackage{times,epsfig}
\usepackage{cite}
\usepackage{subfigure}
\usepackage{graphicx}
\usepackage[cmex10]{amsmath}
\usepackage{url}
\usepackage{multirow}
\usepackage{epstopdf}
\usepackage{algorithm}
\usepackage{algpseudocode}
\usepackage{amsmath}

\usepackage{colortbl}
\usepackage{algorithmicx}
\usepackage{pgfplots}
\pgfplotsset{compat=1.5}
\usepgfplotslibrary{groupplots}
\usepackage{hyperref}
\pgfplotsset{compat=1.5}
\usepgfplotslibrary{groupplots}
\usepackage{booktabs}
\usepackage{threeparttable}
\usepackage{tcolorbox}

\begin{document}
\title{Object Segmentation-Assisted Inter Prediction\\for Versatile Video Coding}	
\author{	
Zhuoyuan Li, \IEEEmembership{Graduate Student Member, IEEE,}
Zikun Yuan, 
Li Li, \IEEEmembership{Member, IEEE,}\\
Dong Liu, \IEEEmembership{Senior Member, IEEE,}
Xiaohu~Tang,~\IEEEmembership{Senior Member, IEEE,}  
and Feng Wu, \IEEEmembership{Fellow, IEEE}\vspace{-1em}
\thanks{
	Date of current version \today. 
	
	\emph{(Z. Li and Z. Yuan contributed equally to this work.)}
		
	\emph{(Corresponding author: Dong Liu.)}
	
	Z. Li, L. Li, D. Liu, and F. Wu are with the CAS Key Laboratory of Technology in Geo-Spatial Information Processing and Application System, University of Science and Technology of China, Hefei 230027, China (e-mail: zhuoyuanli@mail.ustc.edu.cn; lil1@ustc.edu.cn; dongeliu@ustc.edu.cn; fengwu@ustc.edu.cn).
	
	Z. Yuan and X. Tang are with the Information Security
	and National Computing Grid Laboratory, Southwest Jiaotong University, Chengdu 610031, China (e-mail: yuanzikun@my.swjtu.edu.cn;
	xhutang@swjtu.edu.cn).	
}
}
		
\markboth{IEEE Transactions on Broadcasting}
{Li \MakeLowercase{\textit{et al.}}: Object Segmentation-Assisted Inter Prediction for VVC}
\maketitle

\begin{abstract}
In modern video coding standards, block-based inter prediction is widely adopted, which brings high compression efficiency. However, in natural videos, there are usually multiple moving objects of arbitrary shapes, resulting in complex motion fields that are difficult to represent compactly. This problem has been tackled by more flexible block partitioning methods in the Versatile Video Coding (VVC) standard, but the more flexible partitions require more overhead bits to signal and still cannot be made arbitrarily shaped. To address this limitation, we propose an object segmentation-assisted inter prediction method (SAIP), where objects in the reference frames are segmented by some advanced technologies. With a proper indication, the object segmentation mask is translated from the reference frame to the current frame as the arbitrary-shaped partition of different regions without any extra signal. Using the segmentation mask, motion compensation is separately performed for different regions, achieving higher prediction accuracy. The segmentation mask is further used to code the motion vectors of different regions more efficiently. Moreover, the segmentation mask is considered in the joint rate-distortion optimization for motion estimation and partition estimation to derive the motion vector of different regions and partition more accurately. The proposed method is implemented into the VVC reference software, VTM version 12.0. Experimental results show that the proposed method achieves up to 1.98\%, 1.14\%, 0.79\%, and on average 0.82\%, 0.49\%, 0.37\% BD-rate reduction for common test sequences, under the Low-delay P, Low-delay B, and Random Access configurations, respectively.
\end{abstract}
\begin{IEEEkeywords}
Inter prediction, motion compensation, motion estimation, motion vector coding, object segmentation, partition estimation, video coding, VVC.
\end{IEEEkeywords}
\IEEEpeerreviewmaketitle
\section{Introduction}
Block-based hybrid coding framework has been widely adopted in the modern video coding standards, including H.265/High Efficiency Video Coding (HEVC)\cite{sullivan2012overview} and the state-of-the-art Versatile Video Coding (VVC)\cite{bross2021overview}. To reduce the temporal redundancy in sequential frames, inter prediction makes a major contribution to these standards and plays a key role in hybrid video coding schemes\cite{zhang1991predictive, lee2017efficient, chien2021motion, yang2021subblock}.

In general, inter prediction is used to predict the current frame from previously coded frames. To improve the coding performance and reduce the implementation difficulty, block-based inter prediction is used to generate the predictive frame. In block-based inter prediction, the current frame is usually divided into non-overlapping rectangular blocks as the basic units to match prediction. For each block, a motion vector (MV) is used to indicate the displacement between the to-be-coded block and the prediction block in the reference frame. In the entire prediction process, the MV of each block is derived by motion estimation (ME). Then the block-based motion compensation (MC) is utilized to generate the prediction blocks by the indication of MV. 

Block-based ME and MC are generally performed on the assumption that the motion of pixels within a block tends to be uniform. This assumption makes the motion field easy to represent with low complexity in hybrid coding framework. However, in natural videos, there are usually multiple moving objects of arbitrary shapes in video frames, leading to complex motion fields. When using the block-based motion information description method, the block-level MV may not reflect the pixel-level actual motion, and significant prediction errors may be incurred. Therefore, obtaining the flexible partition for moving objects and compactly representing their motion fields is the key to reducing the prediction errors, thereby achieving a good trade-off between the bit consumption of motion information and prediction accuracy. During the development of video coding standards, many partitioning methods have been proposed to approach the efficient representation of complex motion fields. These methods can be classified into three categories: rectangular partition methods, line-based geometric partition methods, and segmentation-based partition methods. 

\begin{figure*}
	\centering
	\vspace{-0.6em}
	\includegraphics[height=45mm]{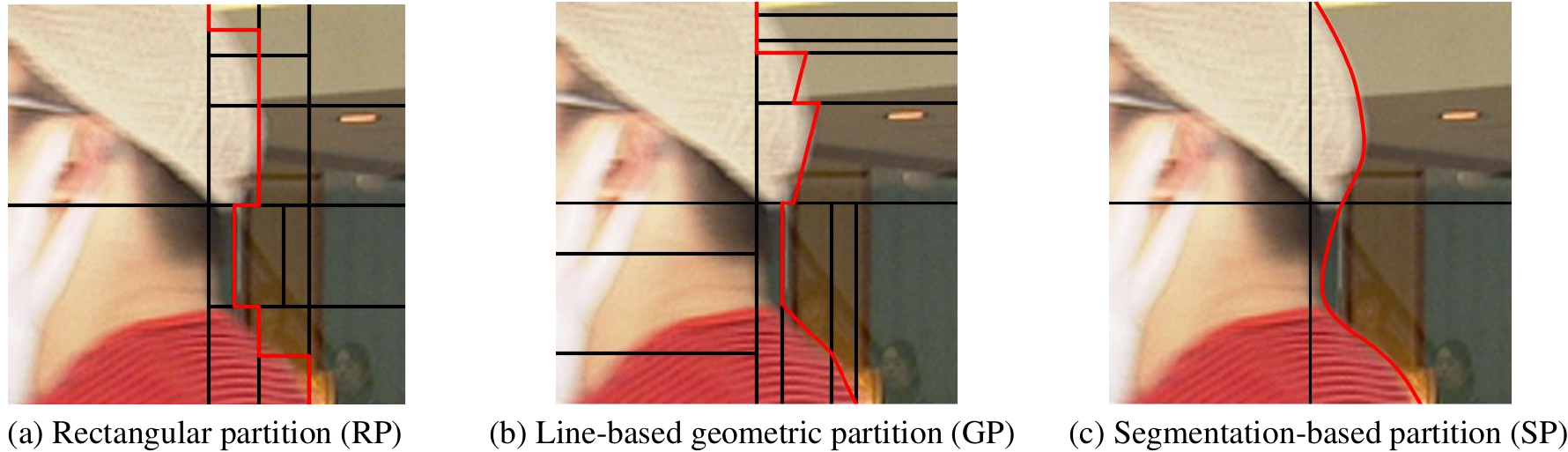}
	\vspace{-0.7em}
	\caption{The partition results of different methods in the actual coding process, where (a) uses the rectangular partition (RP), (b) uses the RP + line-based geometric partition (GP), (c) uses the RP + segmentation-based partition (SP). The block is in the 71-th frame of the \emph{BQMall} sequence.
	}
	\vspace{-1.3em}
	\label{fig:introduction-crop}
\end{figure*}

For rectangular partition (RP) methods, as shown in Fig.~\ref{fig:introduction-crop}\,(a), the rectangular partition structure\cite{wiegand2003overview, kim2012block, huang2021block} is widely used in block-based video coding standards. Partitioning blocks into smaller ones can handle complex motions to some extent. However, due to the inherent shape limitation of the block-based partition, it is difficult to align with natural objects of arbitrary shape. In particular, the block containing object boundary usually consists of regions in different motion fields. If the rectangular partition is used to capture the real motion of each region, it may cost a number of bits.

Line-based geometric partition (GP) methods, as shown in Fig.~\ref{fig:introduction-crop}\,(b), have been intensively studied to capture the non-uniform motion field in HEVC \cite{blaser2017geometry} and VVC \cite{gao2020geometric}. In GP, a predefined straight line is used to split a rectangular block into two wedge-shaped sub-regions. Each sub-region has its motion information and performs MC to obtain its prediction. Although the geometric partition can divide the motion-inconsistent sub-regions in a block to individually perform MC, the straight line can only approximate the discontinuities of the motion field. Meanwhile, extra bit consumption for partition line representation is unavoidable.

Segmentation-based partition (SP) methods, as shown in Fig.~\ref{fig:introduction-crop}\,(c), tentatively introduce the segmentation mask as the partition representation into the existing video coding standards. \cite{chen2007object,blaser2016segmentation,wang2019three} proposed the irregular segmented line to accurately approximate the arbitrary object shape by applying the object-oriented segmentation algorithm on the reference frame/block. The same segmentation algorithm is executed at the encoder and decoder to reduce the overhead bits for the transmission of partition information. Although the related work of SP achieves good results to some extent, they still have two drawbacks. First, previous segmentation algorithms mostly rely on hand-crafted constraints, and it is difficult to approximate the accurate partition of different motion-inconsistent regions. Second, with the segmentation obtained, previous SP methods do not sufficiently take advantage of the object-aware partition information hints for coding process, which limits the potential of fine-grained partition to improve the coding efficiency. 

Recently, with the repaid development of segmentation technologies\cite{minaee2021image, kirillov2023segment, zhou2022survey, cheng2023segment}, moving objects can be pixel-accurately segmented in video frames. Inspired by them, we rethink the assistance of segmentation applied in inter prediction, and sufficiently explore the guidance of accurate object-aware partition information for prediction and coding process in codec. In this paper, we propose an object segmentation-assisted inter prediction (SAIP) method that does not restrict the partitioning shape and further takes advantage of the segmentation information to assist the entire inter prediction process, including MC, motion vector coding (MVC), and ME.

Specifically, in our method, some advanced segmentation technologies are used to segment objects in the reference frames. With a proper indication, the object segmentation mask is translated from the reference block to the coding block as the arbitrary-shaped partition of motion-inconsistent regions. Note that the partition information can be inferred based on the coded motion information instead of being explicitly signaled. With the assistance of the translated segmentation mask, integer-pixel and fractional-pixel MC are separately performed for different regions to achieve higher prediction accuracy. In addition, the segmentation mask is utilized to assist the MVC of different regions by adaptively selecting and ranking the high-probability candidates to construct the candidate list, which considers different qualities of the potential MV candidates to improve the efficiency of motion data transmission. In particular, the segmentation mask is applied to the joint rate-distortion optimization (RDO) for accurate partition estimation and ME at the encoder. The derivation of different MV candidates in ME implicitly considers the various partition candidates in partition estimation for the coding block to estimate jointly.

In summary, we have made the following contributions that will be detailed in this paper:
\begin{itemize}
	\item We propose an object segmentation-assisted inter prediction framework to leverage the segmentation to assist the entire inter prediction process, including MC, MVC, and ME. In this framework, we test the recently advanced segmentation method and verify its effectiveness for video coding.
	\item We design a segmentation-assisted motion compensation (SA-MC) method to separately compensate for the different motion-inconsistent regions with specific design of different MC steps to achieve accurate region-level MC.	
	\item We design a segmentation-assisted motion vector coding (SA-MVC) method to efficiently code the MVs of different motion-inconsistent regions. 
	\item We design a segmentation-assisted rate-distortion optimization (SA-RDO) to joint the partition estimation and ME, and further precisely estimate the partition of coding block and the MV of different motion-inconsistent regions. 
\end{itemize}

\vspace{-0.8em}
\section{Related Work}
\vspace{-0.1em}
In this section, we review the previous work that relates to our research in two aspects. First, we introduce some methods for tackling the representation of complex motion fields in video coding standards. This problem can be viewed as a partition problem, and these methods can be further divided into three categories: rectangular partitioning schemes, line-based geometric partitioning schemes, and segmentation-based partitioning schemes. Second, we introduce some advanced segmentation technologies.

\vspace{-1.1em}
\subsection{Partitioning Schemes}
\vspace{-0.2em}
\subsubsection{Rectangular Partitioning Schemes}
Dividing blocks into smaller ones helps the codec achieve a more accurate motion description. Rectangular partitioning structures are widely used in H.265/HEVC and H.266/VVC, which significantly influence the compression performance. In HEVC, a quadtree (QT) partition structure\cite{kim2012block} is adopted to replace the macro-block used in previous H.264/AVC\cite{wiegand2003overview}. Using the quadtree structure is beneficial for adapting to various local content in natural videos. In VVC, a quadtree plus nested multi-type tree (QTMTT) structure is further introduced, which enables binary-tree (BT), and ternary-tree (TT) partitions besides QT partition \cite{huang2021block}. The QTMTT structure possesses more flexibility because it can generate coding units with more shapes and sizes. Although the coding performance benefits from the flexibility of partition structure, it is difficult to represent the partition of motion area for natural objects with arbitrary shapes. Particularly, the blocks with object boundaries contain multiple sub-regions moving in different directions. With the block-based motion assumption, rectangular partitioning is utilized to capture the real motion characteristics for each region, which will cause more overhead bits. 

\subsubsection{Line-Based Geometric Partitioning Schemes}
Line-based geometric partitioning schemes (GEO) have been intensively studied to capture the non-uniform motion field \cite{blaser2017geometry, gao2016integrated, gao2020geometric, meng2021spatio} during the development of HEVC and VVC. In GEO\cite{gao2016integrated, gao2020geometric}, a predefined straight line is used to split the rectangular block into two wedge-shaped sub-parts. Each sub-part has its MV to describe the motion field and performs MC to obtain the prediction. In VVC, merge-mode-based GEO with 64 partitioning modes has been adopted \cite{gao2020geometric}. The mode is selected for a coding unit from the predefined 64 partition line candidates, and the description information of partition line is represented in the form of the offset parameter $\rho$ and angle parameter $\varphi$. For motion information of sub-parts, only unidirectional motion information is used, and the GEO merge list is derivated from the regular merge list by the parity of the GEO merge index. To economize the bits consumption of GEO, \cite{meng2021spatio} proposed to utilize the spatio-temporal correlation to guide the efficient partitioning mode derivation and motion information coding. Although these methods \cite{blaser2017geometry, gao2016integrated, gao2020geometric, meng2021spatio} can divide the motion-inconsistent sub-parts and represent their motion information with limited bits consumption in a block to perform MC, the straight line can only approximate the discontinuities of the motion field. Meanwhile, bit consumption for partition line representation is unavoidable.

\subsubsection{Segmentation-Based Partitioning Schemes}
To represent the motion field of arbitrary-shaped object  accurately in natural videos, segmentation-based partitioning and coding schemes are tentatively introduced into video coding frameworks and standards \cite{salembier1997segmentation, meier1999video, zheng2011ce2, blaser2018ce10}. In the segmentation-based partitioning schemes, the partition is mainly derived from the reference frame/block by applying a segmentation algorithm at the encoder/decoder. 

In \cite{pardas1996partition}, Pardas \emph{et al}. proposed a segmentation-based partition tree theory for low-bit coding. The partition of multiple regions is derived from the projected merging and segmentation. However, the region-based partition is a rough estimation for the high-texture regions, and how to efficiently extract and represent the irregular partition line information needs to be further studied. In \cite{orchard1993predictive}, the predictive motion field segmentation method was proposed to tackle the problem of partition signaling. To reduce the overhead bits for partition information, \cite{chen2007object, blaser2016segmentation} proposed that the object-oriented partition information was derived from the reference frame by the conventional segmentation method, and one MV was used to signal the position where the segmentation matched the object edges. In \cite{kim2008motion}, Kim \emph{et al}. proposed a novel segmentation method, in which two MVs were signaled and the differences between these two prediction blocks were used for segmentation. Although these methods \cite{chen2007object, blaser2016segmentation, kim2008motion} adopt different strategies to derivate and signal the partition information, the overhead bits are unavoidable. Meanwhile, for ME of different regions, most methods are directly inherited from the normal prediction modes (like \emph{AMVP} and \emph{Merge}) without any modification, and more overhead bits for motion information coding are needed.

Based on these limits, in \cite{wang2019three}, Wang \emph{et al}. proposed a three-zone segmentation-based partitioning method to achieve object-oriented MC for prediction refinement. Based on zone partitioning, different MC methods are used for various zones. However, to reduce the time complexity in partition derivation, \cite{wang2019three} derivated the segmentation of the optimal prediction block obtained by all inter prediction modes 
as the partition of current block. And the motion information of each zone is derivated under the strong assumption that the foreground zone's motion is intense and the background zone's motion is relatively slight. It may not essentially tackle the problem of multiple motion information matching for different zones in a block. And the hypothesis-based partition and MV derivation are individually performed, which limits the accuracy of partition and MV, thereby affecting the prediction quality.

\vspace{-0.8em}
\subsection{Advanced Segmentation Technologies} 
\vspace{-0.1em}

\subsubsection{Image Instance Segmentation (IIS)}
Image segmentation is an essential component in many visual understanding systems, which aims to detect all objects in the input image and assign a pixel-level mask with a category label for each instance of interest in the image\cite{minaee2021image}. It involves partitioning images into multiple segments or objects. Over the past few years, deep learning models have yielded a new generation of image segmentation methods with remarkable performance improvements on popular benchmarks. As the most classic and effective IIS scheme, Mask R-CNN\cite{he2017mask} introduced a fully convolutional mask head to the Faster R-CNN\cite{ren2016faster} detector. Cascade Mask R-CNN\cite{cai2019cascade} combined Cascade R-CNN with Mask R-CNN to achieve good segmentation results. Recently, the Sample Consistency Network (SCNet)\cite{vu2021scnet} proposed a novel training method and reinforced the reciprocal relationships among subtasks, whose average precision (AP) for box and mask prediction was 48.3 and 42.7, respectively. The running time is 38\% faster than the Cascade Mask R-CNN.

\begin{figure*}
	\centering
	\vspace{-1.5em}
	\includegraphics[width=173mm]{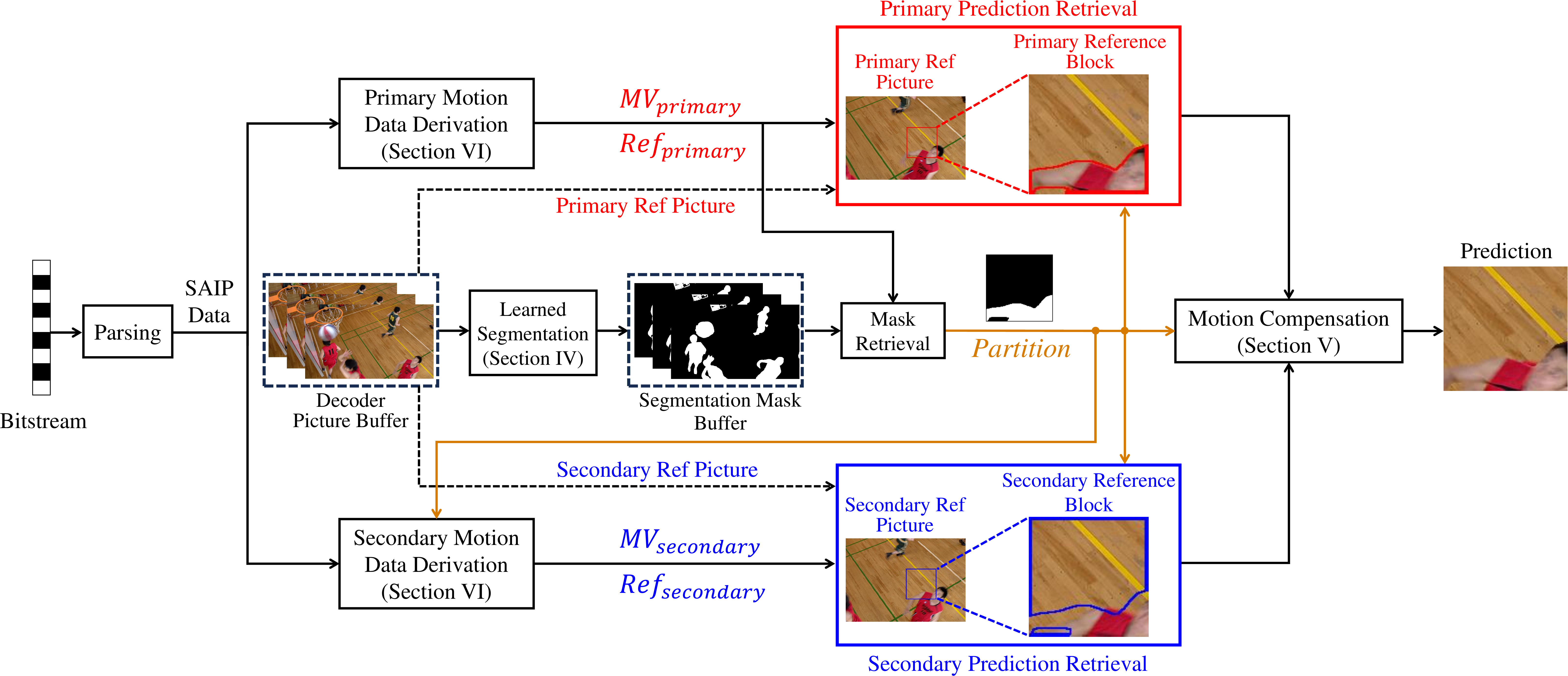}
	\vspace{-0.8em}
	\caption{Illustration of the proposed object segmentation-assisted inter prediction (SAIP) framework from the perspective of decoding. Boxes represent the sub-modules of SAIP, and arrows indicate the data flow direction. Section indicates the corresponding chapter of the module.}
	\label{fig:decoding_process-crop}
	\vspace{-1.5em}
\end{figure*}

\subsubsection{Video Object Segmentation (VOS)}
Video Object Segmentation is an emerging high-level video processing task and has been extensively studied in recent years. It refers to identifying and segmenting the dominant, general objects in video sequences. It is widely used in video analysis and editing-related application scenarios. Recently, most methods fit a model using the initial segmentation\cite{maninis2018video} or leverage temporal propagation\cite{yang2022decoupling, xu2023integrating}, particularly with spatio-temporal matching\cite{hu2018videomatch}. Space-time memory-based networks (STM)\cite{oh2019video} are popular due to their high performance and simplicity. As the best-performing VOS in recent years, space-time correspondence network (STCN)\cite{cheng2021rethinking} subtracts from STM to arrive at a minimalistic form of matching networks and achieves new state-of-the-art results on the benchmark\cite{perazzi2016benchmark}.

\vspace{-0.3em}
\section{Proposed Framework}
\vspace{-0.1em}
Figure~\ref{fig:decoding_process-crop} shows our proposed object segmentation-assisted inter prediction (SAIP) framework from the perspective of decoding. In SAIP, the key lies in that the segmentation information is utilized to assist the overall prediction and coding process in codec. It mainly consists of four modules, including MC (Section \uppercase\expandafter{\romannumeral5}), MVC (Section \uppercase\expandafter{\romannumeral6}), rate-distortion optimization for ME and partition estimation (Section \uppercase\expandafter{\romannumeral7}), segmentation for reference pictures (Section \uppercase\expandafter{\romannumeral4}). The difference between the encoder and decoder is only the additional process of ME and partition estimation. In this section, we briefly introduce the process of SAIP in the decoder.

In the decoding process (Fig.~\ref{fig:decoding_process-crop}\,), different from the previous inter prediction methods, the coding block is divided into multiple motion-inconsistent regions (primary region, secondary region) to match the prediction in SAIP. First, the motion data of the current coding block is obtained by parsing the bitstream. Concerning the motion data for the coding block, two MVs (MV$_{primary}$, MV$_{secondary}$) and reference picture indexes (Ref$_{primary}$, Ref$_{secondary}$) correspond to the motion of different regions (primary motion, secondary motion). The partition of the coding block is translated from the segmentation mask of the reference block by the guidance of MV$_{primary}$. Second, the segmentation mask is utilized to derive the motion information of the secondary region (MV$_{secondary}$). Third, with the assistance of the segmentation mask, the final prediction can be derived by MC based on MVs of different regions.

\section{Learned Segmentation for Reference Pictures}

Obtaining the accurate partition of the motion area is the key to achieving efficient inter prediction. In the traditional coding process, the block-based partition is difficult to align with the irregular shape of the natural object. With the maturity of deep learning-based segmentation methods, the pixel-level demarcation representation ability described by the segmentation map is also increasing. Therefore, the combination of the segmentation method with pixel-level demarcation representation precision and the codec with block-level precision has the potential to more accurately characterize video motions along the time, thus further improving the efficiency of video coding.

Video object segmentation (VOS) aims to identify and segment target instances in a video sequence. According to the strong temporal correlation between multiple video frames, the VOS model mainly propagates the object target information of the label frame to infer the segmentation for the remaining frames. For each frame, the mask represents the segmentation information of the foreground and background. In our work, VOS is utilized to assist the partition process in obtaining the accurate partition of every object's shape. We introduce the recent state-of-the-art segmentation method \cite{vu2021scnet} \cite{cheng2021rethinking} into codec to derive the object segmentation of the multiple reconstructed reference frames and translate the segmentation map of the reference frame to the current frame as the partition of motion-inconsistent regions. Here we focus on the details of how to embed the segmentation module into the codec, and the detailed network design follows the configurations in \cite{vu2021scnet} \cite{cheng2021rethinking}.

The VOS methods can be grouped into two categories in common scenarios: unsupervised and semi-supervised. The difference is that the latter needs to obtain the annotation of the first frame as the tracking label source before the segmentation process. Due to the coding process, the segmentation process needs to be fully automatic. Although some automatic VOS methods can be automated, their segmented precision is far less accurate than that of semi-supervised VOS (SVOS). Therefore, we apply the state-of-the-art SVOS model space-time correspondence network (STCN)\cite{cheng2021rethinking} to our framework. The STCN is a hybrid propagation-based and detection-based method based on a spatio-temporal attention mechanism, which models the complex dynamic spatio-temporal dependencies of instances in multiple frames and excellently captures the motion area to achieve accurate segmentation of instances in each frame. To replace the process of manually extracting the first frame annotations in SVOS with automatic network generation, we propose cooperation between SVOS and image instance segmentation (IIS). IIS is utilized to obtain the annotations required for the first frame of SVOS in the coding process. For IIS, we apply the state-of-the-art IIS model sample consistency network (SCNet)\cite{vu2021scnet}.

\begin{figure}
	\centering
	\vspace{-0.5em}
	\includegraphics[width=80mm]{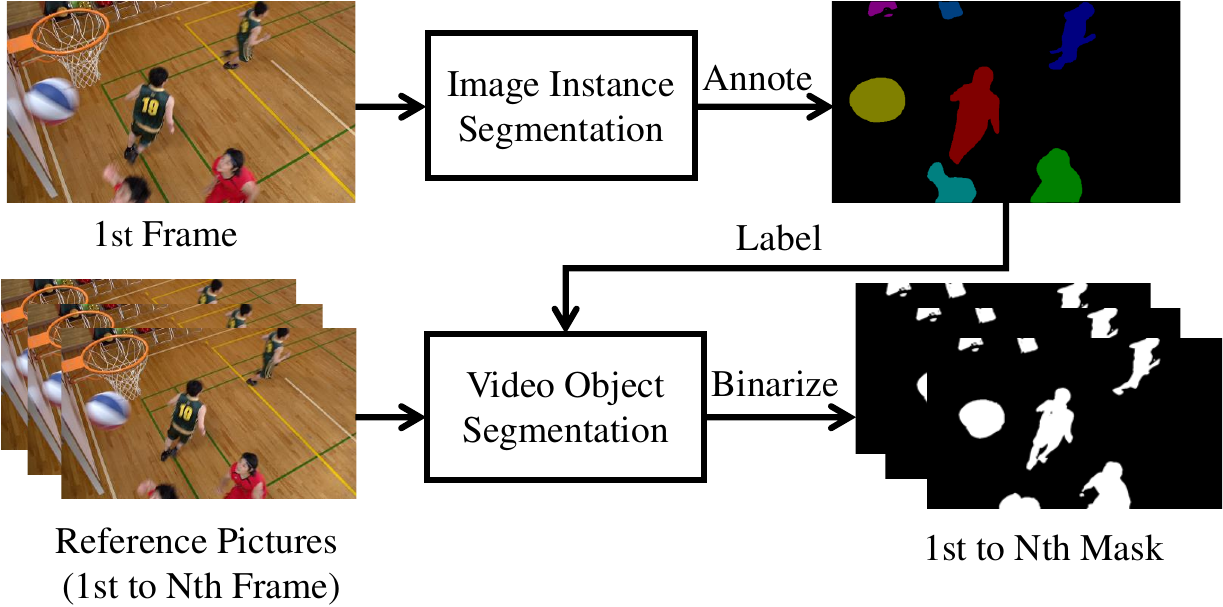}
	\vspace{-0.8em}
	\caption{The object segmentation workflow on the reference frames.}
	\label{fig:segfw}
	\vspace{-1.5em}
\end{figure} 

The derivation process of the reference picture's segmentation map is shown in Fig.~\ref{fig:segfw}\,. From the experiment for different segmentation configurations in the reference pictures, we find that the segmentation accuracy of IIS is influenced by the quantization parameters (QP) of the reference picture, and the VOS is not much affected by the picture quality. To fit the coding frame structure (group of pictures, GOP), SCNet is utilized to segment the first frame of the GOP as the tracking label source of SVOS and the STCN segments the other frames of the GOP.

When dealing with multi-frame segmentation in a GOP, first, SCNet is utilized to detect and segment foreground instances in the first frame of a GOP and obtain the infographic of instances, which contains the bounding box and segmentation of instances. Second, the instance segmentation information of the first frame $S_{0}$ is used as the reference for STCN inference. According to $S_{0}$, the STCN tracks the instance in each subsequent frame and obtains $S_{i}$ of the subsequent frame. To represent different instances in the mask, the segmentation of each frame $S_{i}$ uses different colors to represent different instances and the mask $M_{i}$ of each frame is derived by binarizing $S_{i}$. Finally, we add a buffer memory for each frame to store all $M_{i}$ to facilitate data interaction. The derivation of $M_{i}$ is expressed as follows:
\vspace{-0.3em}
\begin{equation}\label{eq}
	M_{i}(x, y)=
	\begin{cases}
		1 & \text{if } S_{i}(x, y) \in \Phi_{I}\\
		0 & \text{otherwise }
	\end{cases}
\end{equation}
where $M_{i}(x, y)$ and $S_{i}(x, y)$ represent the pixel values of $M_{i}$ and $S_{i}$, and $\Phi_{I}$ denotes the instance sets.

In the case of coding requirements, sequences are also long-term. VOS methods are especially suitable for processing short-term sequences. Although the STCN has largely solved the common challenges of VOS such as object occlusion, fast motion and deformation to a certain extent, STCN still cannot solve more complex motion very well in long-term sequences. We find many scene changes in long-term sequences with new objects appearing and camera movement. Since the STCN does not have the segmentation information of these new objects in the annotation of the first frame, it cannot segment such new objects in time, resulting in a decrease in the accuracy of the segmentation area and contour of the frame. To solve this problem, the interval of the label mask's update can be adjusted (such as 1st-GOP, 2nd-GOP, 3rd-GOP) to adapt to the scene change of different videos. At the beginning of every segmentation interval, SCNet re-derives the object information and re-propagates the mask with the STCN, which is equivalent to updating the object information regularly to guide the accurate segmentation of SVOS. 

\vspace{-0.2em}
\section{Segmentation-Assisted Motion Compensation}

\begin{figure*}
	\centering
	\includegraphics[width=180mm]{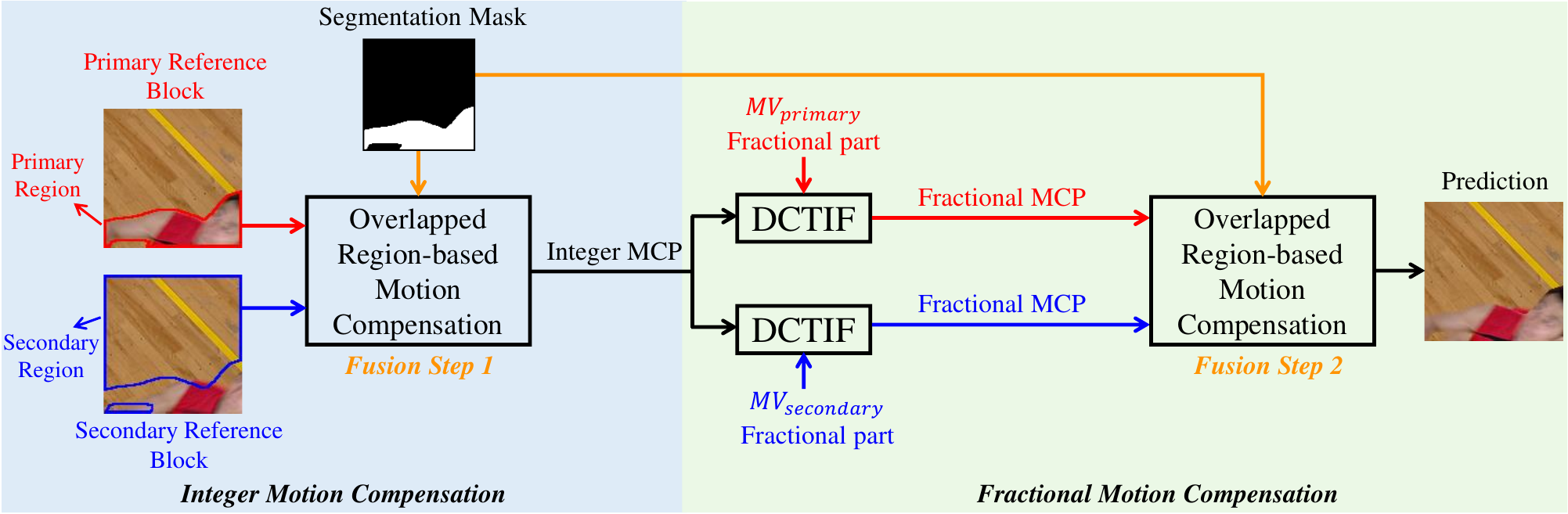}
	\vspace{-0.8em}
	\caption{Schematic representation of the segmentation-assisted motion compensation, including integer and fractional parts.}
	\label{fig:ORMC-framework-crop}
	\vspace{-0.7em}
\end{figure*}

As mentioned in Section \uppercase\expandafter{\romannumeral4}, the reference frame can be segmented with the same setup in the encoder and decoder. The segmentation mask of the reference frame is generated when the coding frame is reconstructed, and stored in buffer memory for segmentation-assisted inter prediction of subsequent frames. With a proper indication, the object segmentation mask of the reference block is translated to the coding block as the arbitrary-shaped partition of motion-inconsistent regions. In this section, we assume that the partition and the motion information of different regions are obtained, and introduce how to utilize the segmentation mask to assist the MC for different motion-inconsistent regions (sub-regions) within the block. 

For different sub-regions, two MVs (MV$_{primary}$ and  MV$_{secondary}$) are used to represent their motion and correspond to two block-level reference blocks named primary prediction and secondary prediction (P$_{primary}$, P$_{secondary}$). Considering the accurate representation of motion information, we propose segmentation-assisted motion compensation (SA-MC) to generate the prediction of different sub-regions within a block, the process of SA-MC is shown in Fig.~\ref{fig:ORMC-framework-crop}\,. In particular, the segmentation mask assists the two steps of MC, including the integer and fractional parts. 

\vspace{-0.8em}
\subsection{Segmentation-Assisted Integer-Pixel Motion Compensation}
\vspace{-0.1em}
In regular integer-pixel MC (IMC), the prediction can be directly copied from the reference frame. In SAIP, if the prediction of each sub-region is directly copied from different reference blocks and joined into a block to generate the block-level prediction, the prediction may result in serious boundary artifacts, thus affecting prediction accuracy. Inspired by overlapped block motion compensation \cite{auyeung1992overlapped, orchard1994overlapped}, we design an overlapped region-based motion compensation (ORMC) method, which uses the arbitrary-shaped (non-linear) partition of motion area and multiple reference blocks to generate the prediction without the region-level boundary artifacts.

\begin{figure}
	\centering
	\includegraphics[width=85mm]{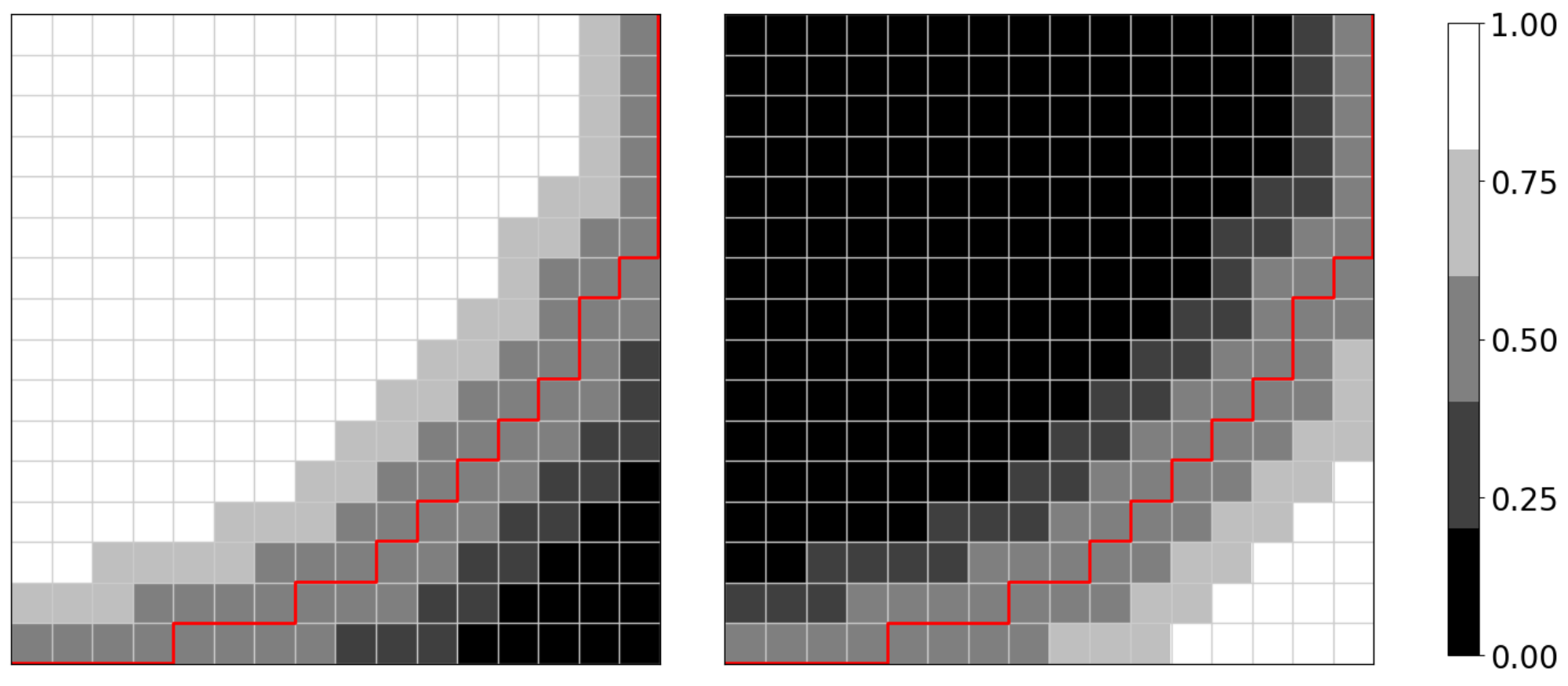}
	\vspace{-0.9em}
	\caption{An example of overlapped region-based motion compensation. The red line is the partition line of motion-inconsistent regions.}
	\vspace{-1.4em}
	\label{fig:ORMC-threezone-crop}
\end{figure}

On both sides of the non-linear partition line, the 2-pixel range region is defined as the edge region. For the edge region to be predicted, not only its own MV but also the MV of the neighboring region can be used to derive the prediction signal. Fig.~\ref{fig:ORMC-threezone-crop} shows a set of ORMC examples. A blending process is applied to the pixels around the non-linear partition line by a distance-based adaptive weighted strategy. Pixels are classified into three categories: one pixel away from the partition line, two pixels away, and more than two pixels away. Each category corresponds to a different weight for the fusion of P$_{primary}$ and P$_{secondary}$ in every pixel.

To classify the pixels around the non-linear partition line, the segmentation mask and edge detection operator are combined to achieve accurate pixel classification. We introduce the Prewitt operator \cite{prewitt1970object, canny1986computational} to classify the pixels of the segmentation mask. First, the first-order differential-based Prewitt operator is introduced to calculate the gradient between the current pixel and adjacent pixels by
\begin{equation}
	\vspace{-0.3em}
	\footnotesize{
		\begin{aligned}{
				\mathbf{G_0}(i,j)=
				\textup{max}\left(\mathbf{S_{0}}\odot\mathbf{P_{0}}, \mathbf{S_{0}}\odot\mathbf{P_{0}}^{T}\right)}
	\end{aligned}}
\end{equation}
where $\mathbf{S_{0}}$ is a 3$\times$3 matrix consisting of the pixel at the $(i, j)$ position in segmentation mask and its surrounding pixels, and $\mathbf{P_{0}}$ is the Prewitt operator:
\vspace{-0.35em}
\begin{equation}
	\footnotesize{
		\mathbf{P_{0}}=\left[
		\begin{array}{ccc}
			-1 & -1 & -1\\
			0 & 0 & 0\\
			1 & 1 & 1\\
		\end{array}\right]
	}
\vspace{-0.35em}
\end{equation}
Then, according to the principle of the operator, the operator is expanded to 5$\times$5 and calculate the gradient of the current pixel and the pixel at a distance of one pixel ($\mathbf{G_{1}}(i,j)$).

Finally, the category $\mathbf{C}$ is derived from the calculation results of $\mathbf{G_0}$ and $\mathbf{G_1}$, by
\begin{equation}\label{eq}
	\footnotesize{
		\mathbf{C}_{i,j}=
		\begin{cases}
			1 & \text{if } |\mathbf{G_0}(i,j)|>0\\
			2 & \text{if } |\mathbf{G_0}(i,j)|=0\ \& \ |\mathbf{G_1}(i,j)|>0\\
			0 & \text{otherwise}
	\end{cases}}
\vspace{-0.2em}
\end{equation}
\vspace{-0.8em}

With the pixel category $\mathbf{C}$ obtained, we allocate different fusion weights for different pixel categories. Therefore, the final prediction $\mathbf{P_{final}}$ can be generated according to the category and the value of the segmentation mask, using different weight values to fuse $\mathbf{P_{primary}}$ and $\mathbf{P_{secondary}}$:
\begin{equation}\label{eq}
	\footnotesize{\mathbf{P_{final}}(i,j)=
		\begin{cases}
			w_{0}\mathbf{P_{primary}}(i,j)+w_{1}\mathbf{P_{secondary}}(i,j) & \text{if } \mathbf{M}_{i,j}=1\\
			w_{1}\mathbf{P_{primary}}(i,j)+w_{0}\mathbf{P_{secondary}}(i,j) & \text{if } \mathbf{M}_{i,j}=0\\
	\end{cases}}
\vspace{-0.1em}
\end{equation}
where the values of $w_0$ and $w_1$ are set by different categories:\vspace{-0.3em}
\begin{equation}\label{eq}
	\footnotesize{
		\begin{cases}
			w_{0}=\tfrac{1}{2},w_{1}=\tfrac{1}{2} & \text{if } \mathbf{C}_{i,j}=1\\
			w_{0}=\tfrac{3}{4},w_{1}=\tfrac{1}{4} & \text{if } \mathbf{C}_{i,j}=2\\
			w_{0}=1,w_{1}=0 & \text{if } \mathbf{C}_{i,j}=0\\
	\end{cases}}
\end{equation}

\vspace{-0.5em}
\subsection{Segmentation-Assisted Fractional-Pixel Motion Compensation}
\vspace{-0.1em}
In regular block-based MC, due to the inherent spatial discretization of digital video, block translation may not happen to be aligned with pixels. Thus, retrieving a block in the reference frames may not predict the coding block well enough. In VVC, discrete cosine transform-based interpolation filter (DCTIF) \cite{lv2012comparison, ugur2013motion} is utilized to generate the fractional-pixel MC prediction (fractional MCP) from the integer-pixel MCP and its adjacent reference pixels. In our case, due to the different motions of sub-regions within a block, the block-level fractional-pixel MC (FMC) method is mismatched for region-level FMC. Fractional-pixel prediction is needed to generate finely for each sub-region. Therefore, considering the combination of sub-regional FMC with the principle of DCTIF, we propose a two-step prediction fusion-based MC process (TWO-STEP), including integer and fractional parts. The key lies in that the integer-pixel prediction fusion is performed in IMC to assist the DCTIF of subsequent FMC. The process of fractional-pixel SA-MC is shown in Fig.~\ref{fig:ORMC-framework-crop}\,.

\begin{figure*}
	\centering
	\vspace{-1.3em}
	\includegraphics[width=177mm]{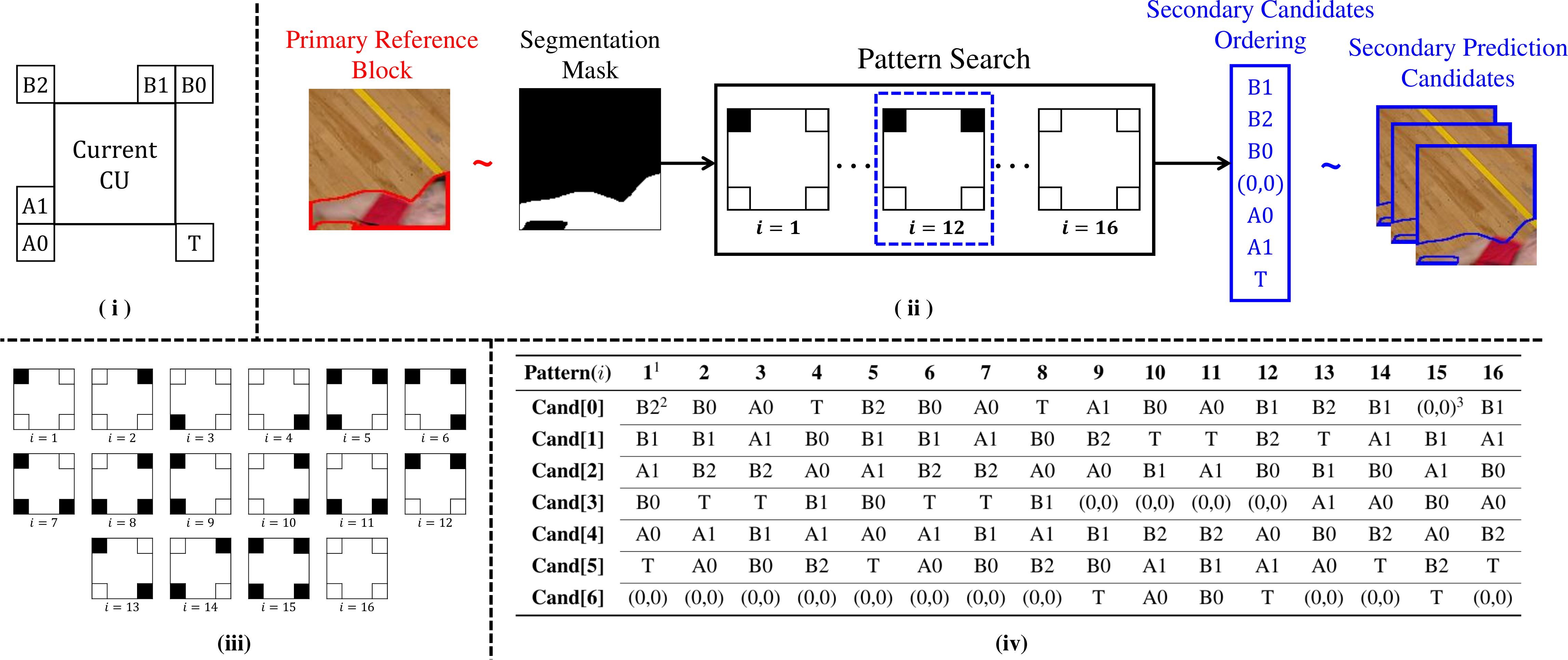}
	\vspace{-0.6em}
	\caption{(i) Position of spatial and temporal merge candidates. (ii) An example of the process of SA-MVC for a coding block. For the pattern search module, the white/black blocks in the corner represent the primary/secondary region; $i$ indicates the pattern ID. (iii) All patterns of motion-inconsistent regions; (iv) Candidate list derivation for secondary region; Top right note 1 indicates the pattern 1 as marked in (iii);  Top right note 2 indicates the neighbor CUs as marked in (i). Top right note 3 indicates zero motion vector (0, 0).}
	\label{fig:SAMVC}
	\vspace{-1.2em}
\end{figure*}
	
First, for the integer-pixel prediction fusion, the integer-pixel MCP is generated from each sub-region's integer-pixel prediction in integer-pixel SA-MC (ORMC), and provides the precise adjacent reference pixels for the interpolation of edge-adjacent fractional pixels in the DCTIF of subsequent FMC. Second, for different sub-region's FMC, the integer-pixel MCP and the fractional part of each sub-region's MV are fed into the DCTIF as the input to generate the fractional-pixel prediction of each sub-region. Finally, with the two fractional-pixel prediction blocks obtained, the fractional-pixel prediction fusion is performed by the same strategy (ORMC) to extract and fuse the fractional MCP of different sub-regions in a block to generate the final fractional-pixel prediction.

\section{Segmentation-Assisted Motion Vector Coding}

In regular inter prediction, block-based inter prediction is widely adopted, and only one MV is used to describe the motion in a coding block. It assumes that the motions within a block tend to be uniform. In regard to complex motion fields, it is difficult to follow the block-based inter mode. When the partition becomes more complicated into more blocks to represent the accurate motion field, the partition and syntax signals will increase the bits consumption. Meanwhile, the precise block partition can only roughly align with natural objects of arbitrary shape, and the real motion field is hard to capture. Thus, flexible partitioning and efficient motion representation are the keys to improving coding performance. Therefore, we propose the segmentation-assisted motion vector coding (SA-MVC) method to use the segmentation mask (object-aware partition information) to guide more accurate motion analysis for fine-grained region level, and further consider different qualities of the potential MV candidates to code the MVs of the different motion-inconsistent regions efficiently. The process of SA-MVC is shown in Fig.~\ref{fig:SAMVC}.

In SAIP, the motion of each coding block is divided into two parts (primary motion and secondary motion) to derive the motion data, which corresponds to two MVs (MV$_{primary}$, MV$_{secondary}$). One is coded by the index of the primary candidate list (length: 71), the list includes regular merge candidates (Merge) and merge with motion vector difference (MMVD) candidates \cite{chien2021motion}. The other one is coded by the index of the secondary candidate list (length: 7), the list is established with the assistance of segmentation mask. 

In the coding process, the primary candidate list is established at first for the coding block, and each candidate can provide a MV (MV$_{primary}$) for the current coding block's rough motion representation. Due to multiple motions in the coding block, MV$_{primary}$ can only consider describing a part of the motion. To further describe the multiple motions, the partition of motion area is needed. By the guidance of MV$_{primary}$, the segmentation mask of the primary reference block can be translated to the current coding block as the partition of the motion area. From the mask, the coding block can be partitioned into two motion-inconsistent regions (primary region, secondary region). For the motion of the primary region, it can be represented by MV$_{primary}$. In order to efficiently represent the motion of other regions, we propose a segmentation-assisted candidate derivation (SA-CD) strategy for the MV candidate derivation of other parts.

\begin{figure*}
	\centering
	\vspace{-1.2em}
	\includegraphics[width=178mm]{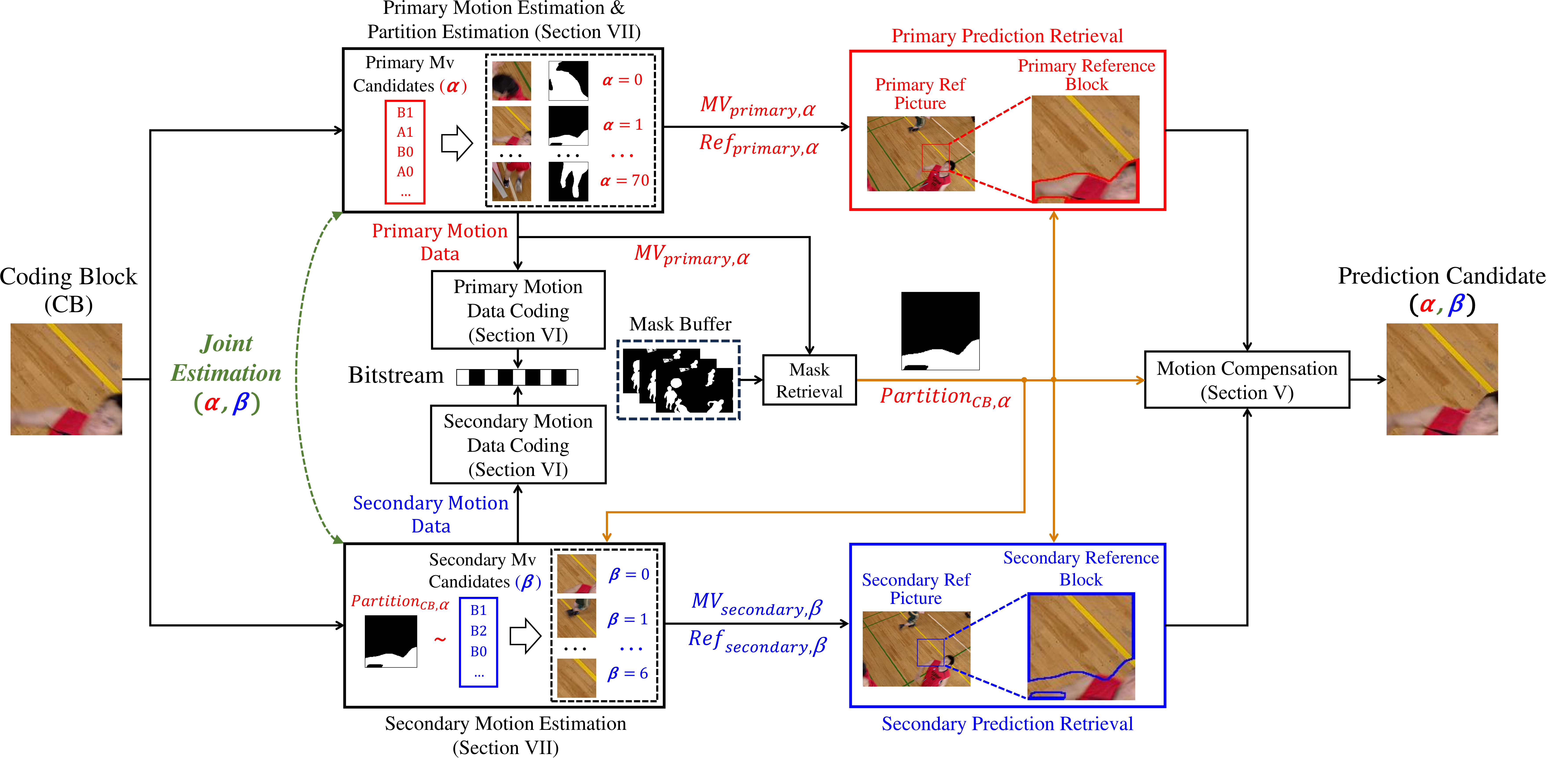}
	\vspace{-0.8em}
	\caption{Illustration of the proposed object segmentation-assisted inter prediction (SAIP) framework with segmentation-assisted rate-distortion optimization (SA-RDO) process (uni-directional case) from the perspective of encoding. Boxes represent the sub-modules of SAIP. Arrows indicate the data flow direction.}
	\label{fig:encoding_process}
	\vspace{-1em}
\end{figure*}

The construction of the regular merge candidate list follows the fixed predefined order to traverse the potential candidates (B1$\rightarrow$A1$\rightarrow$B0$\rightarrow$A0$\rightarrow$B2$\rightarrow$T$\rightarrow$...) to fill the candidate list. Figure~\ref{fig:SAMVC}\,.(i) shows the position of spatial and temporal candidates for the current coding unit (CU), which is used to copy the motion information of the neighboring block for reference. In SA-CD, the segmentation mask (object-aware partition) is utilized as the criterion to perform motion analysis for coding block and further adaptively select and rank the high-probability candidates to construct the candidate list, thereby considering different qualities of the potential MV candidates. The example is shown in Fig.~\ref{fig:SAMVC}\,.(ii). First, the distribution pattern of the secondary region in the partition of the coding block's motion-inconsistent regions is searched. The nearest four-pixel values near the corner are taken as the classification rule. In that case, the four pixels in the upper left and upper right corner of the segmentation mask are all zero (black). It is considered that these corners of the segmentation mask are in the secondary region. Similarly, the four corners can be in the secondary or the primary region, so the distribution of the motion-inconsistent regions can be classified into 16 patterns, all patterns are shown in Fig.~\ref{fig:SAMVC}\,.(iii). Different patterns of distribution of motion-inconsistent regions correspond to different candidate lists. The secondary candidate list is also derived from the candidates in Fig.~\ref{fig:SAMVC}\,.(i). The construction order is determined by the distribution pattern of secondary region, which assumes that the motion information of neighboring blocks closer to the secondary region is more likely to match the motion information of secondary region. Therefore, according to the distribution pattern (blue dotted box) of Fig.~\ref{fig:SAMVC}\,.(ii), the order can be determined as B1$\rightarrow$B2$\rightarrow$B0$\rightarrow$(0,0)$\rightarrow$A0$\rightarrow$A1$\rightarrow$T. The candidate list of all distribution patterns is also shown in Fig.~\ref{fig:SAMVC}\,.(iv).

The motion information of the coding block also affects the coding of the subsequent coding block's ME/MC. In VVC, the motion information is stored in the cache with 4$\times$4 unit. The stored data is used for the MV prediction of the subsequent coding blocks. To assist the MVC of the subsequent coding blocks, the motion information storage of SAIP is adaptively stored based on the partition of the motion area. The value of the segmentation mask determines the storage type of motion information (primary or secondary) for each 4$\times$4 unit.

\section{Segmentation-Assisted Rate-Distortion Optimization for Motion Estimation \\ and Partition Estimation}

For the previous inter prediction methods of VVC, the rate-distortion optimization (RDO) for block-level partition and ME are independent of each other. In our proposed method, due to the region-level partition is introduced to assist the fine-grained inter prediction, ME for different motion-inconsistent regions and region-level partition estimation for coding block are all considered in the RDO process.

For the ME of different motion-inconsistent regions, the accuracy of motion representation depends on whether the partition (segmentation mask) of the motion area can approximate the actual motion field in the coding block. In the previous related work (GEO\cite{gao2020geometric}, PCS-16\cite{wang2019three}, TIP-19\cite{blaser2016segmentation}), template-based or hypothesis-based partition derivation methods are utilized, the fixed search range of partition varieties and individual RDO process (for ME and partition search) limit the partition search space of coding block and the accuracy of MVs, thereby further impacting on the prediction quality. To solve these limits, segmentation-assisted rate-distortion optimization (SA-RDO) is applied to the proposed SAIP framework. In Fig.~\ref{fig:encoding_process}\,, the object segmentation-assisted inter prediction framework with proposed SA-RDO (uni-directional case) is shown from the perspective of encoding. The motion candidates of MV$_{primary}$ imply various partition candidates are used to suppose the motion area partition of current coding block (partition information does not need to be explicitly signaled). And the partition estimation and the ME of primary motion and secondary motion are joint to search the best combination in the whole RDO process. Here we detail the SA-RDO from the perspective of uni-directional and bi-directional prediction.

\vspace{-0.5em}
\subsection{Rate-Distortion Optimization for Uni-directional Prediction}
In codec, the SAIP mode is added as a regular prediction mode in the mode decision list for each coding unit. For a coding block (CB) to try the uni-directional SAIP mode, first, the uni-directional primary candidate list is established. To provide the various partition candidates for accurate partition estimation, the primary candidate list is designed to be longer than the secondary candidate list. Each primary candidate contains MV$_{primary, \alpha}$, Ref$_{primary, \alpha}$ and Partition$_{CB,\alpha}$ with index $\alpha \in \{0 . . . 70\}$. Meanwhile, each primary candidate indicates a different reference block, and the different segmentation mask (representing the different regions by 1/0-value) can be translated to suppose the different motion area partition of current coding block (Partition$_{CB,\alpha}$). For the motion representation of each MV$_{primary, \alpha}$ in the coding block, it is hard to determine whether the MV$_{primary, \alpha}$ represents the 1-value region or the 0-value region's motion. Therefore, the reverse index $j$ $\in \{0, 1\}$ is added to identify the distribution of primary region in the segmentation mask. When the reverse index is 0, the 1-value indicates the primary region and the 0-value indicates the other parts. When the reverse index is 1, the opposite is true. Once the partition and primary region are fixed, the other region corresponds to the secondary region. Second, the uni-directional secondary candidate list is constructed according to SA-CD and Partition$_{CB,\alpha}$, the MV$_{secondary, \beta}$ and Ref$_{secondary, \beta}$ with index $\beta \in \{0 \dots 6\}$ can be selected to represent the motion of other parts in the coding block.

With the uni-directional primary and secondary candidates constructed, the best combination of these candidates is determined by two stages.

\subsubsection{Stage 1} 
The combined uni-directional primary and secondary candidates are used to conduct the SA-MC to generate the prediction of the coding block. The sum of the absolute transformed differences (SATD) of luma between the prediction and the original signal is computed as ${\rm SATD}_{\alpha, j, \beta}$. The rate-distortion (RD) cost $J_{\alpha, j, \beta}$ for $\alpha$, $j$ and $\beta$ can be sorted by
\begin{equation}\label{eq}
	J_{\alpha, j, \beta}={\rm SATD}_{\alpha, j, \beta}+\lambda(R_{\alpha}+R_{j}+R_{\beta}),	
\end{equation}
where $R_{\alpha}$, $R_{j}$ and $R_{\beta}$ denote the estimated rates for the primary candidate index, reverse index, and secondary candidate index, respectively.
\subsubsection{Stage 2} 
From stage 1, the best four combined candidates further apply the residual transform coding and CABAC-based rate estimation to calculate the accurate rate cost $R_{\alpha, j, \beta}$. The distortion over three components between these candidates and the original signal is measured by the sum of squared differences (SSD) as ${\rm SSD}_{\alpha, j, \beta}$. Finally, the optimal $\alpha$, $j$, and $\beta$ can be selected by
\begin{equation}\label{eq}
	J_{\alpha, j, \beta}={\rm SSD}_{\alpha, j, \beta}+\lambda R_{\alpha, j, \beta}.	
\end{equation}

For the uni-directional prediction process of each coding block, 71 (the length of primary candidate list) $\times$ 7 (the length of secondary candidate list) $\times$ 2 (reverse test) = 994 times are calculated and compared in Stage 1 of SA-RDO, and 4 (the length of combined candidate list) $\times$ 2 (residual coding, use or skip) = 8 times are calculated and compared in Stage 2 of SA-RDO, 1002 times maximum calculation in total.

\vspace{-0.5em}
\subsection{Rate-Distortion Optimization for Bi-directional Prediction}

In the bi-directional case of SA-RDO, each bi-directional primary candidate contains MV$_{primary, \alpha, f}$, Ref$_{primary, \alpha, f}$, Partition$_{CB,\alpha, f}$, and MV$_{primary, \alpha, b}$, Ref$_{primary, \alpha, b}$, Partition$_{CB,\alpha, b}$ with index $\alpha \in \{0 . . . 70\}$ . The $f$ indicates the forward reference direction, and $b$ indicates the backward reference direction. For different reference directions, the motion information of the corresponding reference direction of primary candidate indicates a reference block (retrieved by MV$_{primary, \alpha, f}$/Ref$_{primary, \alpha, f}$ or MV$_{primary, \alpha, b}$/Ref$_{primary, \alpha, b}$), and a segmentation mask (Partition$_{CB,\alpha, f}$ or Partition$_{CB,\alpha, b}$) can be translated to suppose the different motion area partition of current coding block in the corresponding reference direction. With the assistance of partition candidate (Partition$_{CB,\alpha, f}$ and Partition$_{CB,\alpha, b}$) in the corresponding reference direction, the subsequent prediction process of each reference direction is the same as the uni-directional case of SA-RDO, and each bi-directional prediction candidate is obtained by the weighted average of different reference direction's prediction.

\begin{figure}
	\centering
	\includegraphics[width=85mm]{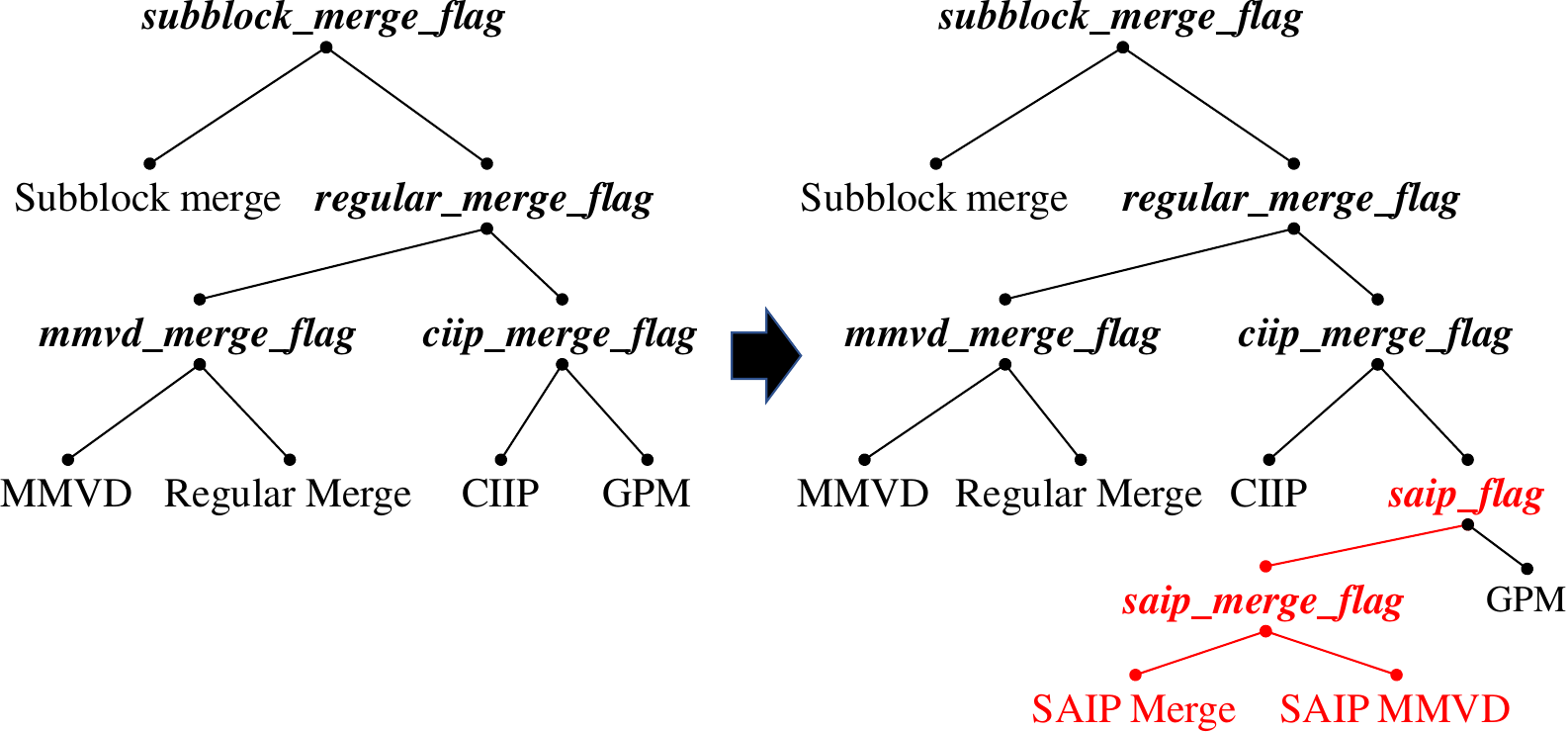}
	\vspace{-0.6em}
	\caption{Entropy coding syntax tree; the left is the original, the right is after adding the SAIP; signaled syntax elements are bold; prediction modes are shown as leaf nodes.}
	\vspace{-0.8em}
	\label{fig:syntax}
\end{figure}

\begin{table}
	\renewcommand\arraystretch{0.7}
	\centering
	\large
	\vspace{-0.2em}
	\caption{SAIP Syntax Elements}
	\vspace{-0.5em}
	\label{tab:Syntax}
	\begin{tabular}{p{13em}lr}
		\toprule  
		\textbf{\small Syntax} & \textbf{\small Descriptor} \\
		\midrule  
		\footnotesize{merge\_data() \{}\\
		\footnotesize{. . .}\\
		\quad\textbf{\textit{\footnotesize{saip\_flag}}} & \footnotesize{ae(v})\\
		\quad\footnotesize{if( saip\_flag )\{}\\
		\quad\textbf{\textit{\footnotesize{saip\_merge\_flag}}} & \footnotesize{ae(v})\\
		\qquad\footnotesize{if( saip\_merge\_flag )\{}\\
		\quad\qquad\textbf{\textit{\footnotesize{saip\_merge\_idx}}} & \footnotesize{ae(v})\\
		\qquad\footnotesize{\}}\\
		\qquad\footnotesize{else\{}\\
		\quad\qquad\textbf{\textit{\footnotesize{saip\_mmvd\_cand\_idx}}} & \footnotesize{ae(v})\\
		\quad\qquad\textbf{\textit{\footnotesize{saip\_mmvd\_distance\_idx}}} & \footnotesize{ae(v})\\
		\quad\qquad\textbf{\textit{\footnotesize{saip\_mmvd\_direction\_idx}}} & \footnotesize{ae(v})\\
		\qquad{\}}\\
		\qquad\textbf{\textit{\footnotesize{saip\_back\_idx}}} & \footnotesize{ae(v})\\
		\qquad\textbf{\textit{\footnotesize{saip\_reverse\_idx}}} & \footnotesize{ae(v})\\
		\quad\footnotesize{\}}\\
		\footnotesize{. . .}\\
		\footnotesize{\}}\\
		\bottomrule 
	\end{tabular}
	\vspace{-0.8em}
\end{table}

\vspace{-0.3em}
\subsection{Complexity Optimization}
To reduce the complexity of SA-RDO, we optimize the encoder algorithm in two aspects, including early termination and parameter pre-calculation. \textit{(1) Early Termination: }The early termination is integrated into the two stages of SA-RDO to reduce the more attempts for the decision of different region's motion data. In \textit{Stage 1}, the minimum cost of each combined candidate is recorded for subsequent candidate's early termination. For each primary candidate, after the rank of SA-CD, the highest-probability (first-order) secondary candidate with the primary candidate is tested and compared with the previous minimum cost. If higher, the attempt of subsequent secondary candidates is terminated. In \textit{Stage 2}, for the best four combined candidates of \textit{Stage 1}, if the cost of the first combined candidate is less than 0.9$\times$ cost of the third combined candidate, the attempt of the last two combined candidates is terminated in \textit{Stage 2}. \textit{(2) Parameter Pre-Calculation:} To prevent the repeated calculation of weighted coefficient in SA-MC, after the reconstructed frame's segmentation mask is generated, the coefficient map is calculated in advance for subsequent frames.

\begin{table*}
	\renewcommand\arraystretch{1.25}
	\centering
	\vspace{-1em}
	\caption{BD-rate Results of our Proposed SAIP Method Compared to VTM-12.0 on CTC Test Sequences}
	\vspace{-0.65em}
	\label{tab:CTC}
	\setlength{\tabcolsep}{1.3mm}
	{
		\begin{tabular}{cclcccccccccccccccc}
			\hline
			\multirow{2}{*}{\textbf{Class}}                                                         & \textbf{Sequence}        &           & \multicolumn{4}{c}{\textbf{Low-delay P (\%)}}            & \multicolumn{1}{l}{}          & \multicolumn{4}{c}{\textbf{Low-delay B (\%)}}            &  & \multicolumn{4}{c}{\textbf{Random Access (\%)}}         \\ \cline{2-2} \cline{4-7} \cline{9-12} \cline{14-17}
			& Name                     &           & Y                & U                & V  &  \textit{Ratio}          &    &          Y                & U                & V    & \textit{Ratio}            &  & Y                & U                & V & \textit{Ratio}                \\ \cline{1-2} \cline{4-7} \cline{9-12} \cline{14-17} 
			\multirow{3}{*}{\textbf{\begin{tabular}[c]{@{}c@{}}\\ClassA1\\ (3840x2160)\end{tabular}}} & \textit{Tango2}          &           & -                & -                & -   & -              &  \multicolumn{1}{l}{}          & -                & -                & -            &  -  &   & -0.21\%          & 0.21\%          & -0.11\%  &   1.19\%      \\
			& \textit{FoodMarket4}     &           & -                & -                & -  & -               & \multicolumn{1}{l}{}          & -                & -                & -            &   - &  & -0.25\%          & -0.25\%           & -0.20\%  & 1.63\%         \\
			& \textit{Campfire}        &           & -                & -                & -   & -             & \multicolumn{1}{l}{}          & -                & -                & -            &   - &   & -0.16\%          & 0.02\%          & -0.12\%  & 1.52\%         \\ 
			
			& \textit{\textbf{Average}}        &           & -                & -                & -   &-              & \multicolumn{1}{l}{}          & -                & -                & -          &  -    &  & \textbf{-0.20\%}          & \textbf{-0.01\%}          & \textbf{-0.14\%}     & \textbf{1.45\%}      \\
			
			\cline{1-2} \cline{4-7} \cline{9-12} \cline{14-17} 
			
			\multirow{3}{*}{\textbf{\begin{tabular}[c]{@{}c@{}}\\ClassA2\\ (3840x2160)\end{tabular}}} & \textit{CatRobot1}       &           & -                & -                & -  &-              & \multicolumn{1}{l}{}          & -                & -                & -         &     -   & & -0.69\%          & -0.01\%           & -0.37\%     & 2.12\%       \\
			& \textit{DaylightRoad2}   &           & -                & -                & -   & -             & \multicolumn{1}{l}{}          & -                & -                & -            &  -  &  & -0.21\%          & -0.37\%          & -0.38\%       & 1.22\%     \\
			& \textit{ParkRunning3}    &           & -                & -                & -    & -            & \multicolumn{1}{l}{}          & -                & -                & -            &   -  &  & -0.06\%          & -0.01\%           & 0.03\%      & 0.72\%      \\
				& \textit{\textbf{Average}}        &           & -                & -                & -     & -           & \multicolumn{1}{l}{}          & -                & -                &  -                &  - & & \textbf{-0.32\%}          & \textbf{-0.13\%}          & \textbf{-0.24\%}      & \textbf{1.35\%}     \\
			\cline{1-2} \cline{4-7} \cline{9-12} \cline{14-17} 
			\multirow{5}{*}{\textbf{\begin{tabular}[c]{@{}c@{}}\\ClassB\\ (1920x1080)\end{tabular}}}  & \textit{MarketPlace}     &           & -0.43\%          & 0.31\%           & -0.62\%  & 2.89\%         &                               & -0.46\%          & -0.78\%          & -0.99\%     &  2.78\%    &  & -0.28\%          & -0.13\%          & -0.07\%      & 1.36\%      \\
			& \textit{RitualDance}     &           & -0.23\%          & -0.35\%          & -0.48\%  & 2.15\%          &                               & -0.28\%          & 0.42\%           & 0.17\%        & 2.27\%   &  & -0.18\%          & 0.27\%          & -0.12\%    & 1.21\%        \\
			& \textit{Cactus}          &           & -0.28\%          & -0.47\%          & -0.38\%   & 2.34\%           &                               & -0.28\%          & -0.24\%          & -0.19\%      & 2.37\%   &  & -0.30\%          & 0.33\%          & -0.40\%    &  1.58\%    \\
			& \textit{BasketballDrive} &           & -0.29\%          & -0.43\%          & -0.66\% & 2.52\%           &                               & -0.25\%          & -0.28\%          & -0.33\%       &  1.69\% &  & -0.20\%          & 0.75\%           & -0.21\%      & 1.45\%     \\
			& \textit{BQTerrace}       &           & -0.03\%           & -0.09\%          & -0.11\%   & 0.73\%          &                               & -0.19\%          & 0.47\%           & 0.49\%         & 1.68\% &  & -0.34\%          & -0.19\%          & -0.40\%     & 1.11\%      \\
				& \textit{\textbf{Average}}        &           & \textbf{-0.25\%}                & \textbf{-0.21\%}                 & \textbf{-0.45\%}    & \textbf{2.13\%}               & \multicolumn{1}{l}{}          & \textbf{-0.30\%}                & \textbf{-0.08\%}                & \textbf{-0.17\%}                & \textbf{2.16\%} &  & \textbf{-0.26\%}          & \textbf{0.21\%}          & \textbf{-0.24\%}     & \textbf{1.34\%}      \\
			\cline{1-2} \cline{4-7} \cline{9-12} \cline{14-17} 
			\multirow{4}{*}{\textbf{\begin{tabular}[c]{@{}c@{}}\\ClassC\\ (832x480)\end{tabular}}}    & \textit{BasketballDrill} &           & -1.83\%          & -1.03\%          & -1.83\%   & 10.73\%         &                               & \underline{-1.14\%}          & -0.74\%          & -1.00\%      &  7.59\%  &  & \underline{-0.79\%}          & -0.08\%          & -0.41\%    & 4.45\%       \\
			& \textit{BQMall}          &           & \underline{-1.98\%}          & -2.84\%          & -2.41\%   & 10.39\%        &                               & -0.84\%          & -1.07\%          & -0.63\%       & 6.54\%  &  & -0.66\%          & -1.08\%           & -0.70\%      & 4.85\%      \\
			& \textit{PartyScene}      &           & -0.27\%          & -0.61\%          & -1.03\%   & 2.92\%         &                               & -0.17\%          & -0.36\%          & -0.16\%       & 2.67\%  &  & -0.17\%          & -0.27\%          & 0.03\%     & 2.29\%       \\
			& \textit{RaceHorsesC}     &           & -0.95\%          & -1.23\%          & -0.39\%   & 5.75\%         &                               & -0.23\%          & -0.13\%           & 0.30\%      &   3.55\% &  & -0.37\%          & -0.40\%           & -0.06\%    & 2.72\%       \\
				& \textit{\textbf{Average}}        &           & \textbf{-1.26\%}                 & \textbf{-1.43\%}                & \textbf{-1.41\%}      & \textbf{7.45\%}           & \multicolumn{1}{l}{}          & \textbf{-0.60\%}                & \textbf{-0.58\%}                & \textbf{-0.37\%}         &    \textbf{5.09\%}   &  & \textbf{-0.50\%}          & \textbf{-0.46\%}          & \textbf{-0.29\%}     & \textbf{3.58\%}      \\
			\cline{1-2} \cline{4-7} \cline{9-12} \cline{14-17} 
			
			\multirow{3}{*}{\textbf{\begin{tabular}[c]{@{}c@{}}\\ClassE\\ (1280x720)\end{tabular}}}   & \textit{FourPeople}      &           & -0.93\%          & -0.64\%           & -0.17\% & 3.56\%           &                               & -0.46\%          & 0.29\%           & -0.36\%     &   2.47\%  &  & -0.41\%                & -0.35\%                & -0.21\%     & 1.37\%            \\
			& \textit{Johnny}          &           & -1.89\%          & -1.29\%          & -1.30\%  & 5.58\%          &                               & -1.12\%        & -2.02\%          & 0.46\%        & 3.54\%  &  & -0.67\%                & -0.25\%                & -0.34\%       & 2.30\%          \\
			& \textit{KristenAndSara}  &           & -0.68\%          & -1.63\%          & 0.02\%   & 3.35\%          &                               & -0.38\%          & -0.80\%          & 0.66\%       &  2.19\%  &  & -0.48\%                 & -0.77\%                & -0.18\%        & 1.36\%        \\
				& \textit{\textbf{Average}}        &           & \textbf{-1.17\%}                & \textbf{-1.19\%}                & \textbf{-0.48\%}          & \textbf{4.16\%}       & \multicolumn{1}{l}{}          & \textbf{-0.66\%}                & \textbf{-0.84\%}                & \textbf{0.25\%}                & \textbf{2.73\%} & & \textbf{-0.52\%}          & \textbf{-0.46\%}          & \textbf{-0.24\%}    & \textbf{1.68\%}       \\
			\cline{1-2} \cline{4-7} \cline{9-12} \cline{14-17} 
			\multicolumn{2}{c}{\textbf{Overall}}                                                                               &           & \textbf{-0.82\%} & \textbf{-0.86\%} & \textbf{-0.78\%} & \textbf{4.58\%}   &  \textbf{}                        & \textbf{-0.49\%} & \textbf{-0.45\%} & \textbf{-0.13\%} & \textbf{3.33\%} &  & \textbf{-0.37\%} & \textbf{-0.14\%} & \textbf{-0.23\%} & \textbf{1.88\%}  \\ \hline
			\multirow{4}{*}{\textbf{\begin{tabular}[c]{@{}c@{}}\\ClassD\\ (416x240)\end{tabular}}}    & \textit{BasketballPass}  &           & -0.61\%          & -0.38\%          & -1.17\%    & 8.58\%       &                               & -0.48\%          & -0.97\%          & -0.74\%     &   5.63\%  &  & -0.44\%          & 0.55\%           & -0.21\%    & 2.78\%      \\
			& \textit{BQSquare}        & \textit{} & -0.78\%          & -2.28\%          & 3.65\%      & 5.82\%      &                               & -0.42\%          & -1.73\%          & 1.76\%     &  4.13\%    &  & -0.53\%          & -0.60\%           & -0.37\%    & 2.68\%        \\
			& \textit{BlowingBubbles}  & \textit{} & -1.02\%          & -1.17\%          & -1.98\%    & 5.85\%       &                               & -0.63\%          & -0.16\%          & 0.39\%       &  4.41\%  &  & -0.50\%          & -0.81\%          & -1.06\%     & 3.40\%      \\
			& \textit{RaceHorses}      & \textit{} & -0.83\%          & -0.21\%          & -0.90\%   & 6.28\%        &                               & -0.10\%          & 0.88\%           & 0.15\%    &   3.68\%   &  & -0.34\%          & 0.94\%           & 0.49\%     & 3.03\%       \\ 
				& \textit{\textbf{Average}}        &           & \textbf{-0.81\%}               & \textbf{-1.01\%}                & \textbf{-0.10\%}    & \textbf{6.63\%}            & \multicolumn{1}{l}{}          & \textbf{-0.41\%}                & \textbf{-0.49\%}                & \textbf{0.39\%}                & \textbf{4.46\%} &  & \textbf{-0.45\%}          & \textbf{0.02\%}          & \textbf{-0.29\%}      & \textbf{2.97\%}     \\ \hline		
		\end{tabular}
	}
	\vspace{-1em}
\end{table*}

\section{Syntax Design}
\vspace{-0.2em}
Syntax elements of SAIP are coded into the bitstream using entropy coding. In the high-level syntax (HLS), two SAIP enabling flags (SAIP\_Merge, SAIP\_MMVD) and the maximum candidate number of the secondary candidate list are coded in the sequence parameter set. The enabling flags are coded with fixed-length code, and the maximum candidate number is coded with the 0-th order exponential-Golomb code.

At the coding unit level, the SAIP syntax elements are signaled inside the merge data syntax. Figure~\ref{fig:syntax} shows the changes in the syntax tree of merge data before and after adding SAIP syntax elements. 
To minimize the impact of the newly increased syntax elements on coding the syntax of other modes, the original syntax tree of merge data has been redesigned. In the new syntax elements tree, saip\_flag is signaled to indicate whether it is in SAIP mode or GPM (GEO) mode, and saip\_merge\_flag is signaled to indicate whether it is in SAIP-Merge mode or SAIP-MMVD mode. 

Table~\ref{tab:Syntax} is the syntax elements table for SAIP. It contains four parts, including the syntax elements of the primary region (Merge: \textit{saip\_merge\_idx}, MMVD: \textit{saip\_mmvd\_cand\_idx}, \textit{saip\_mmvd\_distance\_idx} and \textit{saip\_mmvd\_direction\_idx}), secondary region (\textit{saip\_back\_idx}), mode flag (\textit{saip\_flag} and \textit{saip\_merge\_flag}), and the reverse index (\textit{saip\_reverse\_idx}). These elements are coded by using the context-based adaptive binary arithmetic coding (CABAC) engine. To code the newly increased syntax elements efficiently, the mode flag, reverse index, and the syntax elements of the secondary region are coded by using a new context model.

\vspace{-0.5em}
\section{Experiments}
\subsection{Experimental Settings}
\vspace{-0.2em}
\subsubsection{Learned Segmentation Inference}
For image instance segmentation, following \cite{vu2021scnet}, we use the pre-trained SCNet model and infer the model on the MMDetection platform \cite{chen2019mmdetection}. During inference, object proposals are progressively refined by box branches at different stages. The final classification score is obtained from the scores of multiple classifiers. To exclude object classification error, only objects whose classification score exceeds the threshold $T\in(0, 1)$ can be segmented. In our experiments, $T$ is set to 0.3, 0.6, and 0.9. For video object segmentation, following \cite{cheng2021rethinking}, the STCN adopts the inference code and pre-trained model\,\footnote{\url{https://github.com/hkchengrex/STCN}}\,, and follows the space-time memory networks of the STM\cite{oh2019video}, which stores the previous frames and masks in the memory bank. When inferring the mask of the current frame, five frames are selected from the memory bank and input with the current frame into the STCN.

\subsubsection{Evaluation Configurations}
To evaluate the performance of SAIP in VVC, SAIP is implemented into the VVC reference software VTM-12.0 \footnote{\url{https://vcgit.hhi.fraunhofer.de/jvet/VVCSoftware_VTM}}. Three main profile configurations (Low-delay P (LDP), Low-delay B (LDB), Random Access (RA)) and test sequences specified in \cite{CTCdocument} are used as the test conditions. Following the common test conditions (CTC)\cite{CTCdocument}, four quantization parameters are tested: 22, 27, 32 and 37. For different quantization parameters, different segmentation configurations of the SCNet and STCN are chosen. For the measure of coding performance, the Bjontegaard Delta bit-rate (BD-rate) \cite{2001Calculation} is used as the objective metric to evaluate it. To further measure the characteristic of the proposed method on different sequences, the usage ratio ($Ratio$) is used to validate the efficiency of mode selection, which is calculated by, $Ratio = {N_{test}}/N_{total}$, where $N_{test}$ indicates the number of coding units coded by the proposed method in P/B-frames, and $N_{total}$ indicates the total number of coding units in P/B-frames of the test sequences.

\subsection{Performance}

In this subsection, we show the overall performance of the proposed object segmentation-assisted inter prediction. First, we show the coding performance and encoding/decoding complexity of the proposed SAIP implemented into the VVC reference software VTM-12.0 under CTC\cite{CTCdocument}. Second, we compare SAIP with previous related methods to verify its superior performance. Finally, we test SAIP on the test sequence of specific scenes to further demonstrate its effectiveness.

\begin{table}
	\renewcommand\arraystretch{1.5}
	\centering
	\caption{Time Complexity of Our Method Compared to VTM Anchor}
	\vspace{-0.6em}
	\label{tab:Time}
	\setlength{\tabcolsep}{0.85mm}
	{
		\begin{tabular}{cccccccccc}
			\hline
			\multirow{2}{*}{\textbf{Class}} &  & \multicolumn{2}{c}{\textbf{Low-delay P}} &  & \multicolumn{2}{c}{\textbf{Low-delay B}} &  & \multicolumn{2}{c}{\textbf{Random Access}} \\ \cline{3-4} \cline{6-7} \cline{9-10} 
			&  & EncT               & DecT               &  & EncT               & DecT               &  & EncT                 & DecT                \\ \cline{1-1} \cline{3-4} \cline{6-7} \cline{9-10} 
			\textbf{A}                      &  & -                  & -                  &  & -                  & -                  &  & 183\%                & 116\%               \\ \cline{1-1} \cline{3-4} \cline{6-7} \cline{9-10} 
			\textbf{B}                      &  & 179\%              & 111\%              &  & 171\%              & 112\%              &  & 180\%                & 112\%               \\ \cline{1-1} \cline{3-4} \cline{6-7} \cline{9-10} 
			\textbf{C}                      &  & 159\%              & 109\%              &  & 147\%              & 113\%              &  & 154\%                & 115\%               \\ \cline{1-1} \cline{3-4} \cline{6-7} \cline{9-10} 
			\textbf{E}                      &  & 178\%              & 113\%               &  & 148\%              & 120\%              &  & 179\%                & 118\%               \\ \cline{1-1} \cline{3-4} \cline{6-7} \cline{9-10} 
			\textbf{Overall}                &  & \textbf{172\%}              & \textbf{111\%}              &  & \textbf{155\%}              & \textbf{115\%}              &  & \textbf{174\%}                & \textbf{115\%}               \\ \hline
			\textbf{D}                      &  & 135\%              & 108\%               &  & 123\%              & 119\%              &  & 116\%                & 117\%               \\ \hline
		\end{tabular}
		\vspace{-1em}
	}
\end{table}

\subsubsection{Overall Performance Under Common Test Conditions}
The R-D performance of the entire segmentation-assisted inter prediction framework on the VVC common test sequences is illustrated in Table~\ref{tab:CTC}\,. Y, U, and V represent the R-D performance gain of the three channels of YUV. We can see that our proposed SAIP can achieve on average, 0.82\%, 0.49\% and 0.37\% (marked by font bold), and achieve up to 1.98\%, 1.14\% and 0.79\% (marked by underline) BD-rate reduction (Y component) with high usage ratio on VTM-12.0 for all sequences under the LDP, LDB, and RA configurations. The experimental results show that the proposed framework performs better for sequences with more moving objects and complex motion fields, such as \textit{BasketballDrill}, \textit{BQMall}, \textit{CatRobot}, and \textit{Marketplace}.

In VVC reference software (VTM-12.0), the technology of inter prediction most relevant to SAIP is the geometric partitioning mode (GEO)\cite{gao2020geometric}. Both ours and GEO aim to achieve a flexible description of the motion field and match the appropriate prediction for different regions in a block. In Table~\ref{tab:CTC}\,, under the LDB and RA configurations, GEO is enabled in the anchor by default. In the comparison with anchor, SAIP achieves good performance with GEO enabled. It can be concluded that SAIP can efficiently deal with some complex motion scenarios that GEO is difficult to address. 

\subsubsection{Complexity}
The encoding/decoding time complexity of our proposed method is illustrated in Table~\ref{tab:Time}\,. EncT and DecT represent the encoding/decoding time with the inference time of segmentation network, the network is inferred in parallel with the syntax parsing. The codec time is tested on CPU, and the model is Intel Core i7-11700 @2.50GHz. The inference time of network is tested on GPU, the models are aligned with the \cite{vu2021scnet, cheng2021rethinking}, and the computational complexity is 119 kMAC/pixel. 

Compared with the anchor, under LDP, LDB and RA, for the decoding time, our proposed method does not obviously increase the time complexity in the decoding process. The slight increase in DecT is mainly due to the inference time of the segmentation network. For the encoding time, the increase in EncT is mainly due to the SA-RDO for the partition estimation of coding block and ME of primary and secondary regions to search accurate motion area partition and the MVs of each region.

\begin{table}
	\renewcommand\arraystretch{1.2}
	\centering
	\fontsize{7pt}{9pt}\selectfont 
	\caption{Comparisons with Some Enhanced Geometric-based Partitioning Methods under Low-delay B and Random Access Configurations. Note that Different Methods are Compared to Different Anchors (The Anchor is Shown in Parentheses, Such as VTM-8.0 for TIP-21, VTM-12.0 for SAIP).}
	\vspace{-0.6em}
	\label{tab:CP_GEO}
	\begin{threeparttable}
		\setlength{\tabcolsep}{0.3mm}
		{
		\begin{tabular}{clccclccc}
			\hline
			\multirow{3}{*}{\textbf{Class}} &  & \multicolumn{3}{c}{\textbf{Low-delay B}}                                                                                                                                                                         &  & \multicolumn{3}{c}{\textbf{Random Access}}                                                                                                                                                                       \\ \cline{3-5} \cline{7-9} 
			&  & \textbf{\begin{tabular}[c]{@{}c@{}}TIP-21\tnote{1}\\ (VTM-8.0)\end{tabular}} & \textbf{\begin{tabular}[c]{@{}c@{}}VTM-12.0\\ (VTM-8.0)\end{tabular}} & \textbf{\begin{tabular}[c]{@{}c@{}}SAIP\tnote{2}\\ (VTM-12.0)\end{tabular}} &  & \textbf{\begin{tabular}[c]{@{}c@{}}TIP-21\tnote{1}\\ (VTM-8.0)\end{tabular}} & \textbf{\begin{tabular}[c]{@{}c@{}}VTM-12.0\\ (VTM-8.0)\end{tabular}} & \textbf{\begin{tabular}[c]{@{}c@{}}SAIP\tnote{2}\\ (VTM-12.0)\end{tabular}} \\ \cline{1-1} \cline{3-5} \cline{7-9} 
			\textbf{A1}                     &  & -                                                                   & -                                                                     & -                                                                  &  & -0.20\%                                                             & -5.69\%                                                               & -0.20\%                                                            \\
			\textbf{A2}                     &  & -                                                                   & -                                                                     & -                                                                  &  & -0.21\%                                                             & -12.08\%                                                              & -0.32\%                                                            \\
			\textbf{B}                      &  & -0.31\%                                                             & -1.17\%                                                               & -0.30\%                                                            &  & -0.19\%                                                             & -6.96\%                                                              & -0.26\%                                                            \\
			\textbf{C}                      &  & -0.61\%                                                             & -1.41\%                                                               & -0.60\%                                                            &  & -0.32\%                                                             & -3.22\%                                                               & -0.50\%                                                            \\
			\textbf{E}                      &  & -0.88\%                                                             & -1.29\%                                                               & -0.66\%                                                            &  & -                                                                   & -8.35\%                                                              & -0.52\%                                                            \\ \hline
			\textbf{Overall}                &  & \textbf{-0.55\%}                                                    & \textbf{-1.28\%}                                                      & \textbf{-0.49\%}                                                   &  & \textbf{-0.23\%}                                                    & \textbf{-7.00\%}                                                     & \textbf{-0.37\%}                                                   \\ \hline
			\textbf{D}                      &  & -                                                                   & -1.02\%                                                               & -0.41\%                                                            &  & -                                                                   & -0.83\%                                                               & -0.45\%                                                            \\ \hline
		\end{tabular}}
	\end{threeparttable}
\begin{threeparttable}
\begin{tablenotes}    
	\fontsize{7pt}{9pt}\selectfont   
	\item[1] These results are cited from \cite{meng2021spatio}. Only partial sequences were reported.          
	\item[2] The performance of SAIP on top of VTM-12.0.
\end{tablenotes}
\end{threeparttable}
\vspace{-2.1em}
\end{table}

\begin{table*}
	\renewcommand\arraystretch{1.3}
	\centering
	\scriptsize
	\vspace{-4em}
	\caption{Comparisons with the Other State-of-The-Art Methods under Low-delay B Configuration. Note that Different Methods are Compared to Different Anchors (The Anchor is Shown in Parentheses, Such as HM-14.0 for PCS-16, VTM-1.0 for TIP-19).}
	\label{tab:sota1}
	\vspace{-0.85em}
	\begin{threeparttable}
		\setlength{\tabcolsep}{1.2mm}
		{
			\begin{tabular}{ccccccccccclccclccc}
				\hline
				\multicolumn{19}{c}{\textbf{Low-delay B Configuration (\%)}}                                                                                                                                                                                                                                                                                                                                                                                                                                                                                                         \\ \hline
				\textbf{Sequence}        &           & \textbf{\begin{tabular}[c]{@{}c@{}}PCS-16\tnote{1}\\ (HM-14.0)\end{tabular}} &           & \multicolumn{3}{c}{\textbf{\begin{tabular}[c]{@{}c@{}}TIP-19\tnote{2}\\ (VTM-1.0)\end{tabular}}} &  & \multicolumn{3}{c}{\textbf{\begin{tabular}[c]{@{}c@{}}Simplified VTM-12.0\tnote{3}\\ (HM-14.0)\end{tabular}}} & \multicolumn{1}{c}{} & \multicolumn{3}{c}{\textbf{\begin{tabular}[c]{@{}c@{}}Simplified VTM-12.0\tnote{3}\\ (VTM-1.0)\end{tabular}}} &  & \multicolumn{3}{c}{\textbf{\begin{tabular}[c]{@{}c@{}}SAIP\tnote{4}\\ (Simplified VTM-12.0\tnote{3} )\end{tabular}}} \\ \cline{1-1} \cline{3-3} \cline{5-7} \cline{9-11} \cline{13-15} \cline{17-19} 
				Name                     &           & Y                                                                   &           & Y                           & U                           & V                           &  & Y                                & U                               & V                               &                      & Y                                & U                               & V                               &  & Y                               & U                              & V                              \\ \cline{1-1} \cline{3-3} \cline{5-7} \cline{9-11} \cline{13-15} \cline{17-19} 
				\textit{MarketPlace}     &           & -                                                                   &           & -                           & -                           & -                           &  & -16.81\%                         & -12.53\%                        & -20.41\%                        &                      & -4.86\%                          & 20.24\%                         & 22.04\%                         &  & -2.38\%                         & -2.09\%                        & -2.38\%                        \\
				\textit{RitualDance}     &           & -                                                                   &           & -                           & -                           & -                           &  & -25.05\%                         & -20.87\%                        & -20.66\%                        &                      & -8.90\%                          & 17.14\%                         & 16.08\%                         &  & -2.15\%                         & -1.98\%                        & -1.91\%                        \\
				\textit{Cactus}          &           & -                                                                   &           & -1.00\%                     & -0.90\%                     & -1.00\%                     &  & -20.36\%                         & -6.45\%                         & -6.35\%                         &                      & -9.88\%                          & 22.29\%                         & 22.37\%                         &  & -0.80\%                         & -0.77\%                        & -0.86\%                        \\
				\textit{BasketballDrive} &           & -0.08\%                                                             &           & -0.70\%                     & -0.80\%                     & -1.10\%                     &  & -24.02\%                         & -21.09\%                        & -21.25\%                        &                      & -9.47\%                          & 27.71\%                         & 23.75\%                         &  & -2.20\%                         & -3.00\%                        & -2.88\%                        \\
				\textit{BQTerrace}       &           & -                                                                   &           & -1.60\%                     & -2.00\%                     & -1.70\%                     &  & -16.10\%                         & -15.34\%                        & -28.19\%                        &                      & -8.08\%                          & 18.61\%                         & 11.56\%                         &  & -1.48\%                         & -1.01\%                        & -0.94\%                        \\
				\textit{Kimono}          &           & -0.14\%                                                             &           & -0.90\%                     & -1.00\%                     & -1.40\%                     &  & -15.01\%                         & 4.40\%                          & 0.68\%                          &                      & -6.15\%                          & 20.14\%                         & 20.14\%                         &  & -1.17\%                         & -1.00\%                        & -0.87\%                        \\
				\textit{ParkScene}       &           & -                                                                   &           & -0.60\%                     & -0.40\%                     & -0.70\%                     &  & -16.60\%                         & -3.55\%                         & -4.14\%                         &                      & -6.89\%                          & 20.26\%                         & 22.02\%                         &  & -0.23\%                         & -0.63\%                        & -0.55\%                        \\
				\textit{BasketballDrill} &           & -                                                                   &           & -1.30\%                     & -1.50\%                     & -1.30\%                     &  & -29.14\%                         & -15.11\%                        & -14.63\%                        &                      & -19.76\%                         & 9.37\%                          & 10.45\%                         &  & -4.34\%                         & -4.27\%                        & -4.91\%                        \\
				\textit{BQMall}          &           & -                                                                   &           & -1.50\%                     & -1.10\%                     & -1.00\%                     &  & -22.74\%                         & -14.37\%                        & -13.82\%                        &                      & -12.03\%                         & 14.86\%                         & 14.43\%                         &  & -4.47\%                         & -5.91\%                        & -5.14\%                        \\
				\textit{PartyScene}      &           & -                                                                   &           & -1.80\%                     & -1.40\%                     & -1.10\%                     &  & -16.82\%                         & 0.69\%                          & -1.07\%                         &                      & -9.07\%                          & 21.47\%                         & 22.33\%                         &  & -1.16\%                         & -1.10\%                        & -1.37\%                        \\
				\textit{RaceHorsesC}     &           & -0.95\%                                                             &           & -0.60\%                     & -0.90\%                     & -0.80\%                     &  & -12.52\%                         & 10.91\%                         & 10.28\%                         &                      & -5.06\%                          & 30.48\%                         & 34.81\%                         &  & -2.21\%                         & -2.97\%                        & -3.63\%                        \\
				\textit{BasketballPass}  &           & -                                                                   &           & -0.80\%                     & -0.80\%                     & -1.00\%                     &  & -17.88\%                         & -10.36\%                        & -6.16\%                         &                      & -7.69\%                          & 23.35\%                         & 24.07\%                         &  & -2.41\%                         & -3.60\%                        & -4.02\%                        \\
				\textit{BQSquare}        & \textit{} & -                                                                   &           & -1.90\%                     & -1.60\%                     & -2.00\%                     &  & -13.55\%                         & -10.62\%                        & -12.38\%                        &                      & -8.34\%                          & 7.68\%                          & 5.87\%                          &  & -1.26\%                         & 0.17\%                        & 1.14\%                         \\
				\textit{BlowingBubbles}  & \textit{} & -0.30\%                                                             &           & -1.70\%                     & -1.40\%                     & -1.60\%                     &  & -12.62\%                         & 6.62\%                          & 2.90\%                          &                      & -5.55\%                          & 24.08\%                         & 24.18\%                         &  & -2.18\%                         & -2.35\%                        & -1.14\%                        \\
				\textit{RaceHorses}      & \textit{} & -                                                                   &           & -1.00\%                     & -1.20\%                     & -1.00\%                     &  & -13.47\%                         & 7.69\%                          & 7.84\%                          &                      & -5.10\%                          & 29.82\%                         & 29.66\%                         &  & -2.11\%                         & -2.81\%                        & -2.30\%                        \\
				\textit{FourPeople}      &           & -                                                                   &           & -0.70\%                     & -0.60\%                     & -1.20\%                     &  & -25.13\%                         & -28.39\%                        & -29.28\%                        &                      & -10.46\%                         & 0.92\%                          & -0.80\%                         &  & -2.23\%                         & -1.48\%                        & -1.86\%                        \\
				\textit{Johnny}          &           & -                                                                   &           & -0.90\%                     & -1.00\%                     & -1.20\%                     &  & -25.81\%                         & -30.38\%                        & -34.35\%                        &                      & -10.25\%                         & 9.35\%                          & 6.11\%                          &  & -3.80\%                         & -3.04\%                        & -2.73\%                        \\
				\textit{KristenAndSara}  &           & -                                                                   &           & -1.10\%                     & -1.30\%                     & -0.90\%                     &  & -23.97\%                         & -26.32\%                        & -28.35\%                        &                      & -8.89\%                          & 6.64\%                          & 5.79\%                          &  & -1.53\%                         & -1.07\%                        & 2.21\%                         \\ \cline{1-1} \cline{3-3} \cline{5-7} \cline{9-11} \cline{13-15} \cline{17-19} 
				\textbf{Average}         &           & \textbf{-0.37\%}                                                    & \textbf{} & \textbf{-1.13\%}            & \textbf{-1.12\%}            & \textbf{-1.13\%}            &  & \textbf{-20.72\%}                & \textbf{-12.74\%}               & \textbf{-15.11\%}               &                      & \textbf{-8.69\%}                 & \textbf{18.02\%}                & \textbf{17.49\%}                &  & \textbf{-2.12\%}                & \textbf{-2.16\%}               & \textbf{-1.90\%}               \\ \cline{1-19} 
			\end{tabular}
		}
	\end{threeparttable}
	\begin{threeparttable}
		\begin{tablenotes} 
			\scriptsize   
			\item[1] These results are cited from \cite{blaser2016segmentation}. Only Y-channel BD-rate of partial sequences was reported. 
			\item[2] These results are cited from \cite{wang2019three}.  Only partial sequences were reported.          
			\item[3] VTM-12.0 with several coding tools turned off (MTS, SBT, LFNST, ISP, MMVD, Affine, SbTMVP, LMChroma, DepQuant, IMV, ALF, CCALF, BCW, BcwFast, BIO, CIIP, Geo, AffineAmvr, LMCS, MRL, MIP, DMVR, SMVD, JointCbCr, PROF, ChromaTS). Note that the turned-off coding tools are all not supported in HM-14.0 \cite{HM14.0} and VTM-1.0 \cite{VTM1.0}, and do not affect the inter technology of HM-14.0 and VTM-1.0.
			\item[4] The performance of SAIP on top of Simplified VTM-12.0.
		\end{tablenotes}
	\end{threeparttable}
	\vspace{-5em}
\end{table*}

\begin{table*}
	\renewcommand\arraystretch{1.2}
	\centering
	\scriptsize
	\vspace{-0.8em}
	\caption{Comparisons with the Other State-of-The-Art Methods under Low-delay P and Random Access Configurations. Note that Different Methods are Compared to Different Anchors (The Anchor is Shown in Parentheses, \\Such as HM-14.0 for PCS-16, VTM-1.0 for TIP-19).}
	\label{tab:sota2}
	\vspace{-0.8em}
	\begin{threeparttable}
		\setlength{\tabcolsep}{1.2mm}
		{
			\begin{tabular}{clclclc|ccclccclccc}
				\hline
				\textbf{}                 &  & \multicolumn{5}{c|}{\textbf{Low-delay P Configuration (\%)}}                                                                                                                                                                                                                         & \multicolumn{11}{c}{\textbf{Random Access Configuration (\%)}}                                                                                                                                                                                                                                                                    \\ \hline
				\textbf{Sequence}         &  & \textbf{\begin{tabular}[c]{@{}c@{}}PCS-16\tnote{1}\\ (HM-14.0)\end{tabular}} &  & \textbf{\begin{tabular}[c]{@{}c@{}}Simplified \\ VTM-12.0\tnote{3}\\ (HM-14.0)\end{tabular}} & \multicolumn{1}{c}{\textbf{}} & \textbf{\begin{tabular}[c]{@{}c@{}}SAIP\tnote{4}\\ (Simplified \\ VTM-12.0\tnote{3} )\end{tabular}} & \multicolumn{3}{c}{\textbf{\begin{tabular}[c]{@{}c@{}}TIP-19\tnote{2}\\ (VTM-1.0)\end{tabular}}} & \multicolumn{1}{c}{} & \multicolumn{3}{c}{\textbf{\begin{tabular}[c]{@{}c@{}}Simplified VTM-12.0\tnote{3}\\ (VTM-1.0)\end{tabular}}} &  & \multicolumn{3}{c}{\textbf{\begin{tabular}[c]{@{}c@{}}SAIP\tnote{4}\\ (Simplified VTM-12.0\tnote{3} )\end{tabular}}} \\ \cline{1-1} \cline{3-3} \cline{5-5} \cline{7-10} \cline{12-14} \cline{16-18} 
				Name                      &  & Y                                                                   &  & Y                                                                                   &                               & Y                                                                                & Y                           & U                           & V                           &                      & Y                                 & U                               & V                              &  & Y                               & U                              & V                              \\ \cline{1-1} \cline{3-3} \cline{5-5} \cline{7-10} \cline{12-14} \cline{16-18} 
				\textit{MarketPlace}      &  & -                                                                   &  & -18.51\%                                                                                   &                               & -2.50\%                                                                                & -                           & -                           & -                           &                      & -13.80\%                          & 0.31\%                          & 0.12\%                         &  & -4.25\%                         & -3.40\%                        & -3.37\%                        \\
				\textit{RitualDance}      &  & -                                                                   &  & -25.78\%                                                                                    &                               & -2.28\%                                                                                & -                           & -                           & -                           &                      & -7.38\%                           & 21.24\%                         & 21.23\%                        &  & -2.23\%                         & -2.13\%                        & -2.00\%                        \\
				\textit{Cactus}           &  & -                                                                   &  & -21.26\%                                                                                    &                               & -1.24\%                                                                                & -0.90\%                     & -0.70\%                     & -0.90\%                     &                      & -15.97\%                          & 1.45\%                          & 4.13\%                         &  & -4.39\%                         & -4.08\%                        & -4.00\%                        \\
				\textit{BasketballDrive}  &  & -0.18\%                                                                   &  & -25.95\%                                                                                    &                               & -1.98\%                                                                                & -0.90\%                     & -0.80\%                     & -1.10\%                     &                      & -13.14\%                          & 13.33\%                         & 11.14\%                        &  & -3.17\%                         & -3.29\%                        & -2.89\%                        \\
				\textit{BQTerrace}        &  & -                                                                   &  & -23.45\%                                                                                    &                               & -0.78\%                                                                                & -1.50\%                     & -1.20\%                     & -1.30\%                     &                      & -23.84\%                          & -8.67\%                         & -10.70\%                       &  & -2.13\%                         & -1.62\%                        & -1.55\%                        \\
				\textit{Kimono}           &  & +0.04\%                                                                   &  & -19.11\%                                                                                    &                               & -0.82\%                                                                                & -1.10\%                     & -1.30\%                     & -0.80\%                     &                      & -7.64\%                           & 14.96\%                         & 13.62\%                        &  & -2.23\%                         & -2.78\%                        & -2.32\%                        \\
				\textit{ParkScene}        &  & -                                                                   &  & -16.45\%                                                                                    &                               & -0.57\%                                                                                & -1.10\%                     & -1.20\%                     & -0.70\%                     &                      & -11.18\%                          & 7.55\%                          & 9.02\%                         &  & -1.71\%                         & -1.52\%                        & -1.21\%                        \\
				\textit{BasketballDrill}  &  & -                                                                   &  & -31.04\%                                                                                    &                               & -3.00\%                                                                                & -1.40\%                     & -2.10\%                     & -1.80\%                     &                      & -12.05\%                          & 12.68\%                         & 10.29\%                        &  & -2.70\%                         & -2.99\%                        & -3.24\%                        \\
				\textit{BQMall}           &  & -                                                                   &  & -24.41\%                                                                                    &                               & -3.69\%                                                                                & -1.60\%                     & -1.40\%                     & -1.60\%                     &                      & -10.11\%                          & 13.54\%                         & 13.47\%                        &  & -4.38\%                         & -4.87\%                        & -5.01\%                        \\
				\textit{PartyScene}       &  & -                                                                   &  & -20.51\%                                                                                    &                               & -1.50\%                                                                                & -1.50\%                     & -1.00\%                     & -0.70\%                     &                      & -5.46\%                           & 17.11\%                         & 17.49\%                        &  & -1.68\%                         & -1.94\%                        & -1.93\%                        \\
				\textit{RaceHorsesC}      &  & -1.05\%                                                                   &  & -14.49\%                                                                                    &                               & -2.60\%                                                                                & -0.70\%                     & -0.50\%                     & -0.40\%                     &                      & -10.85\%                          & 18.34\%                         & 19.93\%                        &  & -2.58\%                         & -4.09\%                        & -4.42\%                        \\
				\textit{BasketballPass}   &  & -                                                                   &  & -17.59\%                                                                                    &                               & -2.79\%                                                                                & -0.90\%                     & -1.10\%                     & -1.00\%                     &                      & -8.09\%                           & 18.51\%                         & 17.10\%                        &  & -1.66\%                         & -2.28\%                        & -2.27\%                        \\
				\textit{BQSquare}         &  & -                                                                   &  & -23.14\%                                                                                    &                               & -2.64\%                                                                                & -1.60\%                     & -2.40\%                     & -2.20\%                     &                      & -4.37\%                           & 5.95\%                          & 7.04\%                         &  & -2.17\%                         & -2.07\%                        & -1.50\%                        \\
				\textit{BlowingBubbles}   &  & -0.17\%                                                                   &  & -14.15\%                                                                                    &                               & -2.27\%                                                                                & -1.80\%                     & -1.90\%                     & -2.10\%                     &                      & -7.52\%                           & 15.82\%                         & 17.49\%                        &  & -2.43\%                         & -2.00\%                        & -2.66\%                        \\
				\textit{RaceHorses}       &  & -                                                        &  & -14.06\%                                                                           &                               & -2.99\%                                                                      & -0.80\%                     & -0.70\%                     & -0.70\%                     &                      & -6.52\%                           & 20.78\%                         & 18.72\%                        &  & -2.05\%                         & -3.04\%                        & -2.75\%                        \\
				\textit{FourPeople}       &  & -                                                                   &  & -26.19\%                                                                                    &                               & -1.26\%                                                                                & -0.80\%                     & -0.60\%                     & -0.90\%                     &                      & -14.32\%                          & -6.64\%                         & -8.85\%                        &  & -1.82\%                         & -1.37\%                        & -1.29\%                        \\
				\textit{Johnny}           &  & -                                                                   &  & -28.65\%                                                                                    &                               & -3.38\%                                                                                & -0.80\%                     & -0.50\%                     & -0.20\%                     &                      & -16.01\%                          & -5.92\%                         & -10.13\%                       &  & -3.41\%                         & -2.80\%                        & -3.03\%                        \\
				\textit{KristenAndSara}   &  & -                                                                   &  & -24.97\%                                                                                    &                               & -1.62\%                                                                                & -1.00\%                     & -0.70\%                     & -0.90\%                     &                      & -14.19\%                          & -4.63\%                         & -5.28\%                        &  & -2.07\%                         & -1.63\%                        & -1.56\%                        \\
				\textit{Tango2}           &  & -                                                                   &  & -                                                                                   &                               & -                                                                                & -0.80\%                     & -1.10\%                     & -1.20\%                     &                      & -11.59\%                          & 4.54\%                          & 3.47\%                         &  & -2.90\%                         & -3.32\%                        & -2.85\%                        \\
				\textit{Drums100}         &  & -                                                                   &  & -                                                                                   &                               & -                                                                                & -1.40\%                     & -1.80\%                     & -1.60\%                     &                      & -17.93\%                          & -2.46\%                         & -1.84\%                        &  & -1.78\%                         & -1.93\%                        & -1.23\%                        \\
				\textit{Campfire}         &  & -                                                                   &  & -                                                                                   &                               & -                                                                                & -1.70\%                     & -2.50\%                     & -3.30\%                     &                      & -20.34\%                          & 9.60\%                          & 8.37\%                         &  & -0.72\%                         & -0.47\%                        & -0.88\%                        \\
				\textit{ToddlerFountain2} &  & -                                                                   &  & -                                                                                   &                               & -                                                                                & -0.80\%                     & -0.90\%                     & -1.30\%                     &                      & -6.43\%                           & 48.16\%                         & 38.69\%                        &  & -0.23\%                         & -1.59\%                        & -0.57\%                        \\
				\textit{TrafficFlow}      &  & -                                                                   &  & -                                                                                   &                               & -                                                                                & -0.90\%                     & -1.10\%                     & -0.90\%                     &                      & -14.30\%                          & 2.16\%                          & 2.74\%                         &  & -1.44\%                         & -1.35\%                        & -1.29\%                        \\
				\textit{DaylightRoad2}    &  & -                                                                   &  & -                                                                                   &                               & -                                                                                & -1.10\%                     & -1.00\%                     & -1.10\%                     &                      & -17.23\%                          & -8.60\%                         & -7.49\%                        &  & -3.36\%                         & -2.47\%                        & -2.54\%                        \\
				\textit{FoodMarket4}      &  & -                                                                   &  & -                                                                                   &                               & -                                                                                & -                           & -                           & -                           &                      & -9.80\%                           & 10.05\%                         & 11.89\%                        &  & -1.83\%                         & -1.81\%                        & -1.21\%                        \\
				\textit{CatRobot}         &  & -                                                                   &  & -                                                                                   &                               & -                                                                                & -                           & -                           & -                           &                      & -15.13\%                          & 0.94\%                          & 1.21\%                         &  & -4.24\%                         & -5.45\%                        & -5.28\%                        \\
				\textit{ParkRunning3}     &  & -                                                                   &  & -                                                                                   &                               & -                                                                                & -                           & -                           & -                           &                      & -17.23\%                          & 9.61\%                          & 12.50\%                        &  & -3.92\%                         & -4.07\%                        & -4.32\%                        \\
				
				\cline{1-1} \cline{3-3} \cline{5-5} \cline{7-10} \cline{12-14} \cline{16-18} 
				\textbf{Average}          &  & \textbf{-0.34\%}                                                    &  & \textbf{-22.14\%}                                                                    &                               & \textbf{-2.10\%}                                                                 & \textbf{-1.14\%}            & \textbf{-1.20\%}            & \textbf{-1.22\%}            &                      & \textbf{-12.31\%}                 & \textbf{8.51\%}                 & \textbf{7.98\%}                &  & \textbf{-2.50\%}                & \textbf{-2.61\%}               & \textbf{-2.49\%}               \\ \cline{1-18} 
		\end{tabular}}
	\end{threeparttable}
	\begin{threeparttable}
		\begin{tablenotes}    
			\footnotesize   
			\item Note: The indication of table notes $^1$, $^2$, $^3$, $^4$ is the same as in Table~\ref{tab:sota1}\,.
		\end{tablenotes}
	\end{threeparttable}
	\vspace{-2.1em}
\end{table*}

\begin{table*}
	\renewcommand\arraystretch{1.22}
	\centering
	\caption{BD-rate Results of Our Proposed SAIP Compared to VTM-12.0 on Specific Scenes.}
	\vspace{-0.65em}
	\label{tab:Spscene}
	\setlength{\tabcolsep}{1.5mm}
	{
		\begin{tabular}{clclccccllllccc}
			\hline
			\multirow{2}{*}{\textbf{Sequence}} &  & \multirow{2}{*}{\textbf{Resolution}} &  & \multicolumn{3}{c}{\textbf{Low-Delay P (\%)}} & \multicolumn{1}{l}{} & \multicolumn{3}{c}{\textbf{Low-Delay B (\%)}}                                        &  & \multicolumn{3}{c}{\textbf{Random Access (\%)}} \\ \cline{5-7} \cline{9-11} \cline{13-15} 
			&  &                                      &  & Y             & U             & V             & \multicolumn{1}{l}{} & \multicolumn{1}{c}{Y}      & \multicolumn{1}{c}{U}      & \multicolumn{1}{c}{V}      &  & Y              & U              & V             \\ \cline{1-1} \cline{3-3} \cline{5-7} \cline{9-11} \cline{13-15} 
			\textit{Carphone}                  &  & \textit{\textbf{QCIF}}                        &  & -1.67\%        & -0.09\%        & 2.57\%        &                      & -1.05\%          & -1.85\%           & 0.63\%            &  & -0.80\%         & -1.33\%         & -0.26\%        \\
			\textit{Grandma}                   &  & \textit{\textbf{QCIF}}                        &  & -1.26\%        & -1.91\%        & -0.72\%        &                      & -1.27\%           & -0.90\%           & -1.41\%           &  & -0.18\%         & -0.56\%         & -0.08\%        \\
			\textit{Deadline}                  &  & \textit{\textbf{CIF}}                         &  & -0.92\%        & -0.81\%        & -1.89\%        &                      & -0.81\%           & -0.44\%           & 0.81\%           &  & -0.67\%         & -0.63\%         & -1.19\%        \\
			\textit{Akiyo}                     &  & \textit{\textbf{CIF}}                         &  & -0.69\%        & 0.05\%        & -0.50\%        &                      & -0.86\%           & -1.46\%           & -1.02\%           &  & -0.01\%         & 0.24\%         & 0.41\%        \\
			\textit{Silent}                    &  & \textit{\textbf{CIF}}                         &  & -2.90\%        & -4.03\%        & -2.78\%        &                      & -1.37\%           & -1.26\%           & -0.62\%           &  & -1.02\%         & -0.45\%         & -0.57\%        \\
			\textit{Bowing}                    &  & \textit{\textbf{CIF}}                         &  & -1.71\%        & -3.15\%        & -1.58\%        &                      & -1.08\%           & -0.48\%          & -1.40\%           &  & -0.72\%         & -0.37\%         & -0.69\%        \\
			\textit{Football}                  &  & \textit{\textbf{CIF}}                         &  & -0.71\%        & 1.39\%        & -2.06\%        &                      & -0.70\%           & -1.24\%           & 0.38\%            &  & -0.63\%         & 0.83\%         & -0.47\%        \\
			\textit{Foreman}                   &  & \textit{\textbf{CIF}}                         &  & -0.85\%        & 0.60\%        & -2.98\%        &                      & -0.52\%           & -1.96\%           & 1.11\%            &  & -0.38\%         & -0.38\%         & -0.80\%        \\
			\textit{Vidyo1}                    &  & \textit{\textbf{720P}}                        &  & -0.75\%        & -0.19\%        & -1.06\%        &                      & -0.81\%           & 0.16\%           & -0.39\%           &  & -0.58\%         & -0.75\%         & -0.51\%        \\
			\textit{Vidyo3}                    &  & \textit{\textbf{720P}}                         &  & -1.27\%        & -1.80\%        & -0.88\%        &                      & -0.80\%           & -1.22\%           & -0.29\%           &  & -0.63\%         & -0.45\%         & -0.33\%        \\
			\textit{Vidyo4}                    &  & \textit{\textbf{720P}}                       &  & -0.84\%        & -0.07\%        & 0.65\%        &                      & -0.55\%           & -0.59\%           & -0.99\%           &  & -0.43\%         & -0.37\%         & -0.19\%        \\
			\textit{Crew}                    &  & \textit{\textbf{720P}}                        &  & -0.73\%        & 1.25\%        &-1.20\%        &                      & -0.59\%        & -1.48\%        &-0.98\%           &  & -0.52\%         & -1.82\%         & 0.51\%        \\
			\textit{Aspen}                     &  & \textit{\textbf{1080P}}                       &  & -0.51\%        & -0.20\%        & -0.37\%        &                      & \multicolumn{1}{c}{-0.42\%} & \multicolumn{1}{c}{-0.53\%} & \multicolumn{1}{c}{0.12\%} &  & -0.36\%         & -0.17\%         & 0.08\%        \\
			\textit{Red\_kayak}                &  & \textit{\textbf{1080P}}                       &  & -0.50\%        & -0.30\%        & 0.86\%        &                      & \multicolumn{1}{c}{-0.51\%} & \multicolumn{1}{c}{-1.86\%} & \multicolumn{1}{c}{0.61\%} &  & -0.47\%         & 0.20\%         & -1.02\%        \\
			\textit{Sunflower}                 &  & \textit{\textbf{1080P}}                       &  & -0.60\%        & -0.92\%        & -0.86\%        &                      & \multicolumn{1}{c}{-0.64\%} & \multicolumn{1}{c}{-0.72\%} & \multicolumn{1}{c}{-0.42\%} &  & -0.49\%         & 0.01\%         & -0.25\%        \\
			\textit{FoodMarket2}                 &  & \textit{\textbf{4K}}                       &  & -0.45\%        & -0.63\%        & -0.21\%        &                      & \multicolumn{1}{c}{-0.47\%} & \multicolumn{1}{c}{-1.13\%} & \multicolumn{1}{c}{0.36\%} &  & -0.51\%         & -0.32\%         & 0.24\%        \\
			\textit{BoxingPractice}            &  & \textit{\textbf{4K}}                          &  & -1.22\%        & -1.72\%        & -0.70\%        &                      & \multicolumn{1}{c}{-0.90\%} & \multicolumn{1}{c}{-1.67\%} & \multicolumn{1}{c}{-1.15\%} &  & -0.78\%         & -1.23\%         & -0.55\%        \\
			\textit{Crosswalk}                 &  & \textit{\textbf{4K}}                          &  & -0.67\%        &0.18\%        & -0.85\%        &                      & \multicolumn{1}{c}{-0.63\%} & \multicolumn{1}{c}{-1.86\%} & \multicolumn{1}{c}{-0.35\%} &  & -0.31\%         & -0.70\%         & -0.71\%        \\
			\textit{Narrator}                  &  & \textit{\textbf{4K}}                          &  & -1.31\%        & -1.68\%        & -2.15\%        &                      & \multicolumn{1}{c}{-1.08\%} & \multicolumn{1}{c}{-1.66\%} & \multicolumn{1}{c}{-1.02\%} &  & -0.34\%         & 0.18\%         & 0.10\%        \\
			\textit{SquareAndTimelapse}        &  & \textit{\textbf{4K}}                          &  & -0.65\%        & -0.03\%        & -0.96\%        &                      & \multicolumn{1}{c}{-0.55\%} & \multicolumn{1}{c}{0.07\%} & \multicolumn{1}{c}{-0.54\%} &  & -0.50\%         & -0.68\%         & -0.47\%        \\ \cline{1-1} \cline{3-3} \cline{5-7} \cline{9-11} \cline{13-15} 
			\textbf{Overall}                   &  & \textbf{-}                           &  & \textbf{-1.01\%}        & \textbf{-0.70\%}        & \textbf{-0.88\%}        &                      & \multicolumn{1}{c}{\textbf{-0.78\%}} & \multicolumn{1}{c}{\textbf{-1.10\%}} & \multicolumn{1}{c}{\textbf{-0.33\%}} &  & \textbf{-0.52\%}         & \textbf{-0.44\%}         & \textbf{-0.34\%}        \\ \hline
		\end{tabular}
		\vspace{-1em}
	}
\end{table*}

\subsubsection{Comparisons with State-of-the-Art Methods}
Table~\ref{tab:sota1} and~\ref{tab:sota2} show the experimental data comparing our proposed SAIP with the previous related work on segmentation-based partition methods \cite{blaser2016segmentation} and \cite{wang2019three}, under LDP, LDB and RA configurations (\cite{blaser2016segmentation} did not report RA configuration, \cite{wang2019three} did not report LDP configuration). These related methods use different anchors. Specifically, \cite{blaser2016segmentation} and \cite{wang2019three} use HM-14.0 and VTM-1.0 as the anchor, respectively. We directly cite the results reported in \cite{blaser2016segmentation} and \cite{wang2019three}, since our reimplementation may not work as well as the authors'. HM-14.0\cite{HM14.0} and VTM-1.0\cite{VTM1.0} do not contain some new technical tools in the latest version of VVC, which have a large performance gap with VTM-12.0. To make a fair comparison with related work, we test SAIP in a simplified VTM-12.0 with some coding tools turned off, and compare it with \cite{blaser2016segmentation} \cite{wang2019three} on HEVC and VVC test sequences. Note that the turned-off coding tools in simplified VTM-12.0 are all not supported in HM-14.0\cite{HM14.0} and VTM-1.0\cite{VTM1.0}. (The number of coded frames of these methods in Table~\ref{tab:sota1} and~\ref{tab:sota2} follows the number of coded frames specified in the Common Test Condition of HEVC\cite{bossen2013common} and VVC\cite{CTCdocument}.) 
From Table ~\ref{tab:sota1} and~\ref{tab:sota2}\,, it is observed that the proposed SAIP achieves a better relative BD-rate reduction on higher version than the other methods on low versions, which demonstrates the superior performance of the proposed method.

In terms of performance comparison with previous segmentation-based partition methods\cite{blaser2016segmentation} \cite{wang2019three}, the bitrate savings of the proposed method mainly come from the following four aspects: (1) We design a complete segmentation-assisted inter prediction framework, which sufficiently explores the assistance effect of segmentation in each sub-modules of inter prediction. In addition, the flexible and accurate motion area partition greatly improves the efficiency of inter prediction for motion-inconsistent regions. (2) For the MC of multiple motion-inconsistent regions in a block, the segmentation-assisted process  can achieve higher prediction accuracy. It leverages the assistance of segmentation for the multi-stage MC. (3) For the MVC of different regions, the proposed SA-CD fully utilizes the segmentation mask to provide suitable prediction candidates efficiently and further reduces the transmission of motion data. (4) For the ME of different regions, the motion information candidates implicitly indicate partition candidates for the partition estimation of coding block, which provide various partition candidates to suppose the actual motion situation and jointly consider the partition estimation and ME in the RDO process to estimate the partition and MVs accurately.

In addition, we also compare our proposed method with some enhanced geometric-based partitioning methods \cite{meng2021spatio}, under LDB and RA configurations. The comparison result is shown in Table~\ref{tab:CP_GEO}\,. 
Specifically, \cite{meng2021spatio} uses VTM-8.0 as the anchor, and we directly cite the results reported in \cite{meng2021spatio} (did not report LDP configuration). From Table~\ref{tab:CP_GEO}\,, it is observed that the proposed SAIP achieves a good relative BD-rate reduction on the VTM higher version than the other enhanced geometric-based partitioning methods on the VTM low version, which further verifies the efficiency of the proposed method.

\subsubsection{On Specific Scenes}
According to the performance of SAIP under CTC and the comparison with other methods, the proposed object-aware prediction framework is beneficial to the scene with moving objects and complex motion fields (such as the walkers of \textit{BQMall} shown in Fig\,.~\ref{fig:partition} and the athletes of \textit {BasketballDrill}), and achieves high performance on these scenes. To further verify the characteristics and advantages of SAIP, we test SAIP on some selected test sequences of specific scenes (diverse scenes, multiple moving objects, complex motions) from previous standards' CTC \cite{sullivan2012overview, wiegand2003overview} and popular datasets \cite{xiph}. Table~\ref{tab:Spscene} shows the performance of partial sequences on these specific scenes.
We can see that our proposed SAIP can achieve up to 2.90\%, 1.37\%, 1.02\%, and on average 1.01\%, 0.78\%, 0.52\% BD-rate reduction (Y component) on VTM-12.0. The results demonstrate the SAIP can achieve accurate object-aware prediction on specific scenes with more moving objects, and further improve the flexibility of motion modeling in VVC.

\begin{table*}
	\renewcommand\arraystretch{1.25}
	\centering  
	\caption{Ablation Study on SA-MC and SA-MVC Based on VTM-12.0}
	\vspace{-0.6em}
	\label{tab:Ab}
	\setlength{\tabcolsep}{1.3mm}
	{
		\begin{tabular}{ccccccccccccccccc}
			\hline
			\multicolumn{17}{c}{\textbf{Low-delay P Configuration (\%)}}                                                                                                                                                                                                                                                             \\ \hline
			\multirow{2}{*}{\textbf{Class}} &           & \multicolumn{3}{c}{\textbf{ONE-STEP SA-MC}}                  &           & \multicolumn{3}{c}{\textbf{TWO-STEP SA-MC}}                  & \textbf{} & \multicolumn{3}{c}{\textbf{w/o SA-MVC}}                 &           & \multicolumn{3}{c}{\textbf{SAIP (SA-MC+SA-MVC)}}         \\ \cline{3-5} \cline{7-9} \cline{11-13} \cline{15-17} 
			&           & Y                & U                & V                &           & Y                & U                & V                &           & Y                & U                & V                &           & Y                & U                & V                \\ \cline{1-1} \cline{3-5} \cline{7-9} \cline{11-13} \cline{15-17} 
			B                               &           & -0.05\%          & 0.03\%          & -0.29\%           &           & -0.10\%          & -0.17\%          & -0.11\%          &           & -0.11\%          & -0.48\%          & -0.20\%           &           & -0.25\%          & -0.21\%          & -0.45\%          \\
			C                               &           & -0.52\%          & -0.64\%          & -0.69\%          &           & -0.89\%          & -1.21\%          & -0.70\%          &           & -1.07\%          & -1.23\%          & -1.31\%          &           & -1.26\%          & -1.43\%          & -1.41\%          \\
			E                               &           & -0.49\%          & -0.51\%          & -0.31\%          &           & -0.68\%          & -0.90\%          & -0.74\%          &           & -0.64\%          & -1.46\%          & -0.48\%          &           & -1.17\%          & -1.19\%          & -0.48\%          \\ \cline{1-1} \cline{3-5} \cline{7-9} \cline{11-13} \cline{15-17} 
			\textbf{Avg.}                   &           & \textbf{-0.32\%} & \textbf{-0.33\%} & \textbf{-0.43\%} & \textbf{} & \textbf{-0.51\%} & \textbf{-0.70\%} & \textbf{-0.46\%} &           & \textbf{-0.56\%} & \textbf{-0.98\%} & \textbf{-0.64\%} &           & \textbf{-0.82\%} & \textbf{-0.86\%} & \textbf{-0.78\%} \\ \hline
			D                               &           & -0.58\%          & -0.52\%          & -0.93\%          &           & -0.83\%          & -0.89\%          & -1.28\%          &           & -0.60\%          & -0.64\%          & -0.64\%          &           & -0.81\%          & -1.01\%          & -0.10\%          \\ \cline{1-1} \cline{3-17} 
			&           &                  &                  &                  &           &                  &                  &                  &           &                  &                  &                  &           &                  &                  &                  \\ 
			
			\vspace{-2em}
			\\
			
			\hline
			\multicolumn{17}{c}{\textbf{Low-delay B Configuration (\%)}}                                                                                                                                                                                                                                                             \\ \hline
			\multirow{2}{*}{\textbf{Class}} &           & \multicolumn{3}{c}{\textbf{ONE-STEP SA-MC}}                  &           & \multicolumn{3}{c}{\textbf{TWO-STEP SA-MC}}                  &           & \multicolumn{3}{c}{\textbf{w/o SA-MVC}}                 & \textbf{} & \multicolumn{3}{c}{\textbf{SAIP (SA-MC+SA-MVC)}}         \\ \cline{3-5} \cline{7-9} \cline{11-13} \cline{15-17} 
			&           & Y                & U                & V                &           & Y                & U                & V                &           & Y                & U                & V                &           & Y                & U                & V                \\ \cline{1-1} \cline{3-5} \cline{7-9} \cline{11-13} \cline{15-17} 
			B                               &           & 0.01\%          & -0.08\%          & 0.14\%           &           & -0.11\%          & -0.13\%          & 0.00\%           &           & -0.12\%          & -0.37\%          & 0.11\%          &           & -0.30\%          & -0.08\%          & -0.17\%          \\
			C                               &           & -0.07\%          & 0.00\%          & -0.29\%          &           & -0.42\%          & -0.79\%          & -0.08\%          &           & -0.32\%          & -0.37\%          & -0.75\%          &           & -0.60\%          & -0.58\%          & -0.37\%          \\
			E                               &           & -0.33\%          & -0.12\%          & -0.37\%          &           & -0.43\%          & -0.62\%           & -0.20\%           &           & -0.27\%          & -0.45\%          & -0.07\%          &           & -0.66\%          & -0.84\%          & 0.25\%           \\ \cline{1-1} \cline{3-5} \cline{7-9} \cline{11-13} \cline{15-17} 
			\textbf{Avg.}                   &           & \textbf{-0.10\%} & \textbf{-0.06\%} & \textbf{-0.13\%}  & \textbf{} & \textbf{-0.29\%} & \textbf{-0.47\%} & \textbf{-0.08\%} &           & \textbf{-0.22\%} & \textbf{-0.39\%} & \textbf{-0.22\%} &           & \textbf{-0.49\%} & \textbf{-0.45\%} & \textbf{-0.13\%} \\ \hline
			D                               &           & -0.37\%          & -0.65\%           & 0.45\%           &           & -0.47\%          & 0.14\%          & 0.35\%           &           & -0.12\%          & 0.21\%          & 0.93\%           &           & -0.41\%          & -0.49\%          & 0.39\%          \\ \hline
			&           &                  &                  &                  &           &                  &                  &                  &           &                  &                  &                  &           &                  &                  &                  \\ 
			
			\vspace{-2em}
			\\
			\hline
			\multicolumn{17}{c}{\textbf{Random Access Configuration (\%)}}                                                                                                                                                                                                                                                           \\ \hline
			\multirow{2}{*}{\textbf{Class}} &           & \multicolumn{3}{c}{\textbf{ONE-STEP SA-MC}}                  & \textbf{} & \multicolumn{3}{c}{\textbf{TWO-STEP SA-MC}}                  & \textbf{} & \multicolumn{3}{c}{\textbf{w/o SA-MVC}}                 & \textbf{} & \multicolumn{3}{c}{\textbf{SAIP (SA-MC+SA-MVC)}}         \\ \cline{3-5} \cline{7-9} \cline{11-13} \cline{15-17} 
			& \textbf{} & Y                & U                & V                &           & Y                & U                & V                &           & Y                & U                & V                &           & Y                & U                & V                \\ \cline{1-1} \cline{3-5} \cline{7-9} \cline{11-13} \cline{15-17} 
			A                               &           & 0.80\%          & 0.31\%          & 0.41\%           &           &  -0.06\%          & -0.13\%          & -0.13\%          &           & -0.08\%          & 0.02\%          & 0.02\%          &           & -0.26\%          & -0.07\%        & -0.19\%         \\
			B                               &           & 0.76\%           & 0.55\%           & 0.65\%           &           & -0.05\%           & 0.07\%           & -0.10\%          &           & -0.09\%          & -0.04\%          & -0.04\%           &           & -0.26\%          & 0.21\%           & -0.24\%          \\
			C                               &           & 0.55\%           & 0.34\%           & 0.63\%           &           & -0.20\%          & -0.24\%          & -0.26\%           &           & -0.18\%          & -0.30\%          & -0.03\%          &           & -0.50\%          & -0.46\%          & -0.29\%           \\
			E                               &           & 0.40\%           & 0.26\%           & 0.29\%           &           & -0.06\%           & 0.16\%           & 0.09\%           &           & -0.13\%           & -0.10\%          & 0.07\%           &           & -0.52\%          & -0.46\%          & -0.24\%           \\ \cline{1-1} \cline{3-5} \cline{7-9} \cline{11-13} \cline{15-17} 
			\textbf{Avg.}                   &           & \textbf{0.67\%}  & \textbf{0.38\%}  & \textbf{0.51\%}  & \textbf{} & \textbf{-0.09\%} & \textbf{-0.05\%}  & \textbf{-0.11\%}  &           & \textbf{-0.11\%} & \textbf{-0.09\%} & \textbf{0.00\%}  &           & \textbf{-0.37\%} & \textbf{-0.14\%} & \textbf{-0.23\%} \\ \hline
			D                               &           & 0.70\%           & 0.67\%           & 0.40\%           &           & -0.12\%          & 0.52\%          & 0.12\%           &           & -0.09\%          & 0.27\%          & -0.06\%          &           & -0.45\%          & 0.02\%           & -0.29\%          \\ \hline
		\end{tabular}		
	}
	\vspace{-0.5em}
\end{table*}

\subsection{Performance Analysis}

In this subsection, we analyze the performance of SAIP through the effectiveness of core modules and the mode selection results of SAIP in detail.

\subsubsection{Ablation Study}
To demonstrate the contributions of two core modules in our scheme, we conduct the ablation experiments on proposed SA-MC and SA-MVC. 

First, we validate the effectiveness of SA-MC and study the influence of different components of SA-MC, including the two-step prediction fusion (TWO-STEP) and overlapped region-based weighted strategy (ORMC). As shown in Table~\ref{tab:Ab}\,, we introduce two variants for comparison with SA-MC. ONE-STEP indicates that the final prediction is directly derived according to the segmentation mask from the block-level primary and secondary predictions at the end of fractional-part MC. TWO-STEP indicates that prediction fusion is performed on the integer and fractional parts of MC. The integer-part fusion mainly aims to generate the accurate integer pixels for the subsequent fractional MC process, and provides the precise adjacent reference pixels for the interpolation of edge-adjacent fractional pixels. SAIP indicates that the TWO-STEP and ORMC are all used in SA-MC. Through the experimental results, we draw two conclusions: (1) For the comparison of ONE-STEP and SAIP, the experimental results show that SA-MC is superior to the direct multi-region MC, which demonstrates the assistance of a segmentation mask is beneficial to generating accurate prediction. (2) For the comparison of two variants and SAIP, the experimental results show that TWO-STEP and ORMC have a significant effect on SA-MC, and further demonstrate that TWO-STEP can better assist DCTIF in generating accurate fractional-pixel prediction and ORMC is beneficial to reducing the boundary artifacts.

Second, we validate the effectiveness of SA-MVC. In Table\,~\ref{tab:Ab}\,, we introduce a variant for comparison with SA-MVC. Instead of using SA-MVC, the secondary candidate list is constructed according to the regular merge candidate order. For the comparison of variant and SAIP, the experimental results show that our proposed MVC method helps achieve more bits saving than the regular method, which  demonstrates the assistance of a segmentation mask is beneficial to coding the MV efficiently.

\vspace{-0.1em}
\subsubsection{Comparison of Mode Selection}
\begin{figure}
	\centering
	\includegraphics[width=85mm]{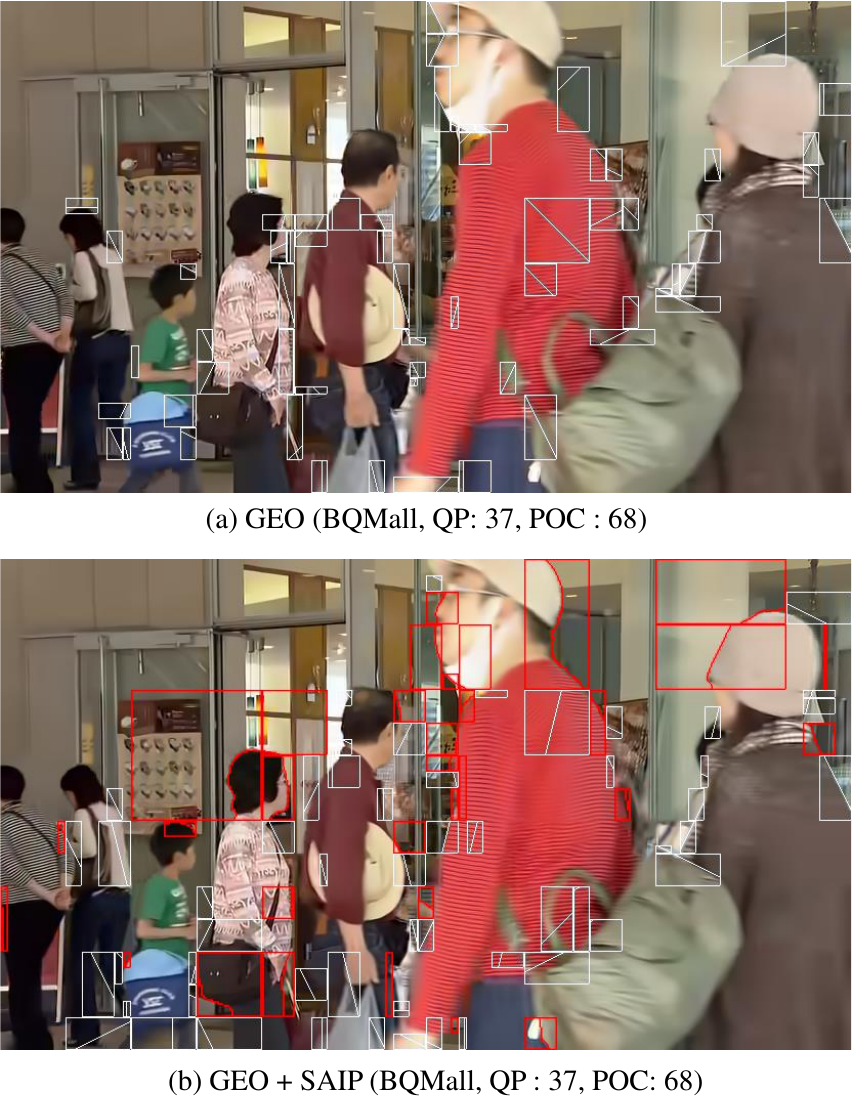}
	\vspace{-0.7em}
	\caption{Mode Selection Results on VTM-12.0 (Configuration: LDB), where (a) uses the GEO mode (anchor: VTM-12.0), (b) uses the GEO mode and proposed SAIP mode. The white/red blocks indicate the block with GEO/SAIP selected.  (More results are shown in the supplementary material.)
	}
	\label{fig:partition}
	\vspace{-1.2em}
\end{figure}

Figure~\ref{fig:partition} shows the mode selected results of the GEO mode and our proposed SAIP mode, more results are shown in Section \uppercase\expandafter{\romannumeral3} of supplementary material. Compared with GEO, SAIP possesses more flexible motion field description capability, which focuses on more motion-inconsistent regions at the edges of moving objects and handles more difficult cases such as the outer outline of the people in \emph{BQMall}. Meanwhile, the SAIP not only tackles the efficient representation of complex motion fields, but also tends to hit the larger block to save the bit consumption of the refined partition.

\vspace{-0.5em}
\subsection{Limitation and Potential}
\vspace{-0.10em}
In this subsection, we discuss the limitations and potential of the proposed SAIP through its effectiveness on different scenes and configurations. 

\begin{table*}
	\renewcommand\arraystretch{2.1}
	\centering
	\scriptsize 
	\caption{BD-rate Results of Our Proposed SAIP Method Compared to VTM-12.0 on the Test Sequences of VOS's Train and Validation Dataset\cite{caelles20182018,xu2018youtube,ding2023mose}, the Detailed Settings of Dataset are Shown in Section \uppercase\expandafter{\romannumeral2}-B of Supplementary Material.}
	\vspace{-1.9em}
	\begin{threeparttable}
		\setlength{\tabcolsep}{0.22mm}
		\label{tab:potential}
		{
			\begin{tabular}{cllcllclclccccccccccccc}
				\hline
				\multicolumn{2}{c}{\multirow{2}{*}{\textbf{Dataset}}} &  & \multicolumn{2}{c}{\multirow{2}{*}{\begin{tabular}[c]{@{}c@{}}Sequence\\ Num\end{tabular}}} & \textbf{} & \multirow{2}{*}{\begin{tabular}[c]{@{}c@{}}Objects\\ Num\tnote{1}\end{tabular}} & \textbf{} & \multirow{2}{*}{\begin{tabular}[c]{@{}c@{}}Selected\\ Character\end{tabular}}                                               &  & \multicolumn{2}{c}{LDB (R\tnote{2} /L\tnote{3} )}          & \multicolumn{1}{l}{\textbf{}} & \multicolumn{2}{c}{RA (R/L)}           & \multicolumn{1}{l}{} & Accuracy (R-L)\tnote{4}        & \multicolumn{1}{l}{} & \multicolumn{2}{c}{\textbf{LDB $\Delta$}\tnote{6}}                & \multicolumn{1}{l}{} & \multicolumn{2}{c}{\textbf{RA $\Delta$}\tnote{6}}                 \\ \cline{11-12} \cline{14-15} \cline{17-17} \cline{19-20} \cline{22-23} 
				\multicolumn{2}{c}{}                                  &  & \multicolumn{2}{c}{}                                                                        &           &                                                                        &           &                                                                                                                             &  & \multicolumn{1}{c|}{Average\tnote{5}\,\,} & Highest\tnote{5}\, & \multicolumn{1}{l}{\textbf{}} & \multicolumn{1}{c|}{Average} & Highest & \multicolumn{1}{l}{} & $\mathcal{J}$ \& $\mathcal{F}$ Mean\tnote{4} & \multicolumn{1}{l}{} & \multicolumn{1}{c|}{Average} & Highest & \multicolumn{1}{l}{} & \multicolumn{1}{c|}{Average} & Highest \\ \cline{1-2} \cline{4-5} \cline{7-7} \cline{9-9} \cline{11-12} \cline{14-15} \cline{17-17} \cline{19-20} \cline{22-23} 
				\multicolumn{2}{c}{\textit{DAVIS-17}\cite{caelles20182018}}                 &  & \multicolumn{2}{c}{30}                                                                      &           & 99                                                                    &           & \multirow{4}{*}{\begin{tabular}[c]{@{}c@{}}Diverse Scenes\\ Multiple Objects\\ Complex Motions\\ High Quality\end{tabular}} &  & \multicolumn{1}{c|}{-0.21\%/-0.51\%}     & -0.36\%/-0.81\%     &                               & \multicolumn{1}{c|}{-0.11\%/-0.39\%}     & -0.33\%/-0.72\%     &                      & 83.2/83.1/81.6/77.7                             &                      & \multicolumn{1}{c|}{-0.30\%}             & -0.45\%             &                      & \multicolumn{1}{c|}{-0.28\%}              & -0.39\%              \\ \cline{1-2} \cline{4-5} \cline{7-7} \cline{11-12} \cline{14-15} \cline{17-17} \cline{19-20} \cline{22-23} 
				\multicolumn{2}{l}{\textit{Youtube-VOS}\cite{xu2018youtube}}              &  & \multicolumn{2}{c}{30}                                                                      &           & 120                                                                     &           &                                                                                                                             &  & \multicolumn{1}{c|}{-0.19\%/-0.41\%}     & -0.31\%/-0.77\%     &                               & \multicolumn{1}{c|}{-0.04\%/-0.31\%}     & -0.22\%/-0.64\%     &                      & 80.7/79.3/77.3/74.9                             &                      & \multicolumn{1}{c|}{-0.22\%}              & -0.46\%              &                      & \multicolumn{1}{c|}{-0.27\%}              & -0.42\%              \\ \cline{1-2} \cline{4-5} \cline{7-7} \cline{11-12} \cline{14-15} \cline{17-17} \cline{19-20} \cline{22-23} 
				\multicolumn{2}{c}{\textit{MOSE}\cite{ding2023mose}}                     &  & \multicolumn{2}{c}{30}                                                                      &           & 295                                                                    &           &                                                                                                                             &  & \multicolumn{1}{c|}{-0.16\%/-0.43\%}     & -0.41\%/-0.73\%     &                               & \multicolumn{1}{c|}{0.09\%/-0.27\%}     & -0.25\%/-0.53\%     &                      & 72.3/68.5/62.7/57.2                             &                      & \multicolumn{1}{c|}{-0.27\%}              & -0.32\%              &                      & \multicolumn{1}{c|}{-0.36\%}              & -0.28\%              \\ \cline{1-2} \cline{4-5} \cline{7-7} \cline{11-12} \cline{14-15} \cline{17-17} \cline{19-20} \cline{22-23} 
				\multicolumn{2}{c}{\textbf{Overall}}                  &  & \multicolumn{2}{c}{\textbf{90}}                                                             & \textbf{} & \textbf{514}                                                           &           &                                                                                                                             &  & \multicolumn{1}{c|}{\textbf{-0.19\%/-0.45\%}}     & \textbf{-0.36\%/-0.77\%}     &                               & \multicolumn{1}{c|}{\textbf{-0.02\%/-0.32\%}}     & \textbf{-0.27\%/-0.63\%}     &                      & \textbf{78.7/77.0/73.9/69.9}                             &                      & \multicolumn{1}{c|}{\textbf{-0.26\%}}              & \textbf{-0.41\%}              &                      & \multicolumn{1}{c|}{\textbf{-0.30\%}}              & \textbf{-0.36\%}              \\ \hline
		\end{tabular}}
		\vspace{0.1em}
	\end{threeparttable}
	\begin{threeparttable}
		\begin{tablenotes}    
			\scriptsize     
			\item \tnote{1} Objects Num denotes the total object number of the test sequences. The object number of each test sequence is shown in Section \uppercase\expandafter{\romannumeral2}-B of Supplementary Material.
			\item \tnote{2} R indicates the test results are performed by the proposed reference frame-based segmentation method (Section \uppercase\expandafter{\romannumeral4}). 
			\item \tnote{3} L indicates the test results are performed by the reference frame's hand-annotated label segmentation mask (open-source annotation of train and validation dataset  \cite{caelles20182018,xu2018youtube,ding2023mose}).
			\item \tnote{4} Accuracy (R-L) indicates the similarity between the proposed segmentation method's result and hand-annotated label mask, which is evaluated by the mean of region similarity $\mathcal{J}$ and contour accuracy $\mathcal{F}$ \cite{perazzi2016benchmark} ($\mathcal{J}_s$ / $\mathcal{F}_s$ in \cite{xu2018youtube}). The four results correspond to the segmentation similarity under different quantization parameters (22, 27, 32, 37).\vspace{0.15em}
			\item \tnote{5} Average indicates the average BD-rate reduction on the Y Component of proposed method on the test sequences. Highest indicates the highest performance among them.
			\item \tnote{6} $\Delta$ indicates the SAIP's performance difference between R- and L-based segmentation generation method.
		\end{tablenotes}
	\end{threeparttable}
	\vspace{-1em}
\end{table*}

\subsubsection{Limitation}
Based on the above experiments, we find that the performance of SAIP is limited by segmentation accuracy. Here we analyze it in two aspects: (1) \textit{Segmentation Technology}: From the CTC results of SAIP (Table~\ref{tab:CTC}\,), we find that our method performs poorly on some sequences with special cases (extreme motion, serious occlusion, large resolution), which is mainly limited by the poor segmentation results, such as \textit{BQTerrence} (camera motion), \textit{Tango2} (indistinguishable foreground and background), \textit{Campfire} (dense moving objects). (These failure segmentation cases are shown in Section \uppercase\expandafter{\romannumeral2}-A of supplementary material.) Meanwhile, learned segmentation schemes have not yet been widely generalized to large-resolution video, such as 1080P, 4k (resolution of Class A, B). The datasets and benchmarks \cite{caelles20182018, xu2018youtube} of these segmentation schemes are mainly evaluated on medium resolution (480P-720P). Due to the limited capability and generalization of the learned segmentation method \cite{vu2021scnet, cheng2021rethinking}, it's difficult to generate accurate segmentation to assist the motion prediction of SAIP in some challenging scenes. (2) \textit{Dependence of Temporal-domain Reference Structure}: From the comparison of SAIP's CTC results under different configurations in Table~\ref{tab:CTC}\,, we find that the reference structure also affects its performance. Due to the motion area partition of each block being derived from the reference block, the motion area partition of the long-distance reference block is difficult to adapt to the motion situation of the current coding block and further influences prediction accuracy, such as the RA reference structure case.

\subsubsection{Potential}
To explore the potential improvement brought by segmentation accuracy, we test the SAIP on the datasets (train and validation datasets of VOS Benchmark \cite{caelles20182018,xu2018youtube,ding2023mose}) with hand-annotated label segmentation mask to avoid the above limitations. For each dataset, we select 30 videos with multiple moving objects, and test the SAIP under different segmentation settings, including the proposed segmentation method (Section \uppercase\expandafter{\romannumeral4}) and hand-annotated label segmentation mask. The detailed settings and experimental results are shown in Table~\ref{tab:potential}\,. Table~\ref{tab:potential} shows the difference of SAIP's performance under different segmentation settings (LDB/RA $\Delta$) can achieve up to -0.41\%, -0.36\%, and on average -0.26\%, -0.30\% BD-rate reduction (Y component) on VTM-12.0. (The visualization of the different settings' segmentation results is shown in Section \uppercase\expandafter{\romannumeral2}-B of supplementary material.)  It demonstrates that the SAIP can achieve higher performance with the assistance of the better segmentation mask, and exhibits positive performance potential with continuous enhancement of advanced segmentation technology \cite{kirillov2023segment, zhou2022survey, cheng2023segment} in the future.

\vspace{-0.3em}
\section{Conclusion}
\vspace{-0.2em}
In this paper, object segmentation-assisted inter prediction is proposed to improve the representation of complex motion fields in inter-frame coding. We introduce  segmentation information to help the codec achieve the flexible partition of motion area, and further utilize it to assist the entire inter prediction process. We exploit the potential semantic information to improve the coding performance under the advanced coding standard. We have conducted extensive experiments to verify the effectiveness of our method. Experimental results show that the proposed method achieves on average 0.82\%, 0.49\% and 0.37\% BD-rate reduction compared to VTM anchor, under the LDP, LDB, RA profiles, respectively, and also performs better than the other state-of-the-art methods.

In future work, we will further extend the SAIP with flexible motion modeling in VVC. The design of the current SAIP mainly addresses translation motion in the coding scene. We may further consider the additional prediction process to solve the more complex situation like rotation and zooming. We will further explore the combination of SAIP and the motion model \cite{watanabe1991windowed, li2017efficient, wang2003temporal, gao2020decoder, li2022global}. In addition, the potential semantic information can also be used to improve the performance of the other modules in the codec, such as quantization\cite{schwarz2021quantization, xu2023complexity}, transform\cite{zhao2021transform, da2021class}, and post-processing\cite{karczewicz2021vvc, chen2015adaptive}. We will further extend the other modules with the assistance of segmentation information to achieve good performance in future video coding.

\vspace{-0.3em}
\bibliographystyle{IEEEtran}
\bibliography{IEEEexample}
\end{document}


\title{\emph{Supplementary Material for}\\Object Segmentation-Assisted Inter Prediction\\for Versatile Video Coding}	
\author{	
	Zhuoyuan Li,
	Zikun Yuan, 
	Li Li,
	Dong Liu,
	Xiaohu~Tang,  
	and Feng Wu}
\maketitle

\section{Learned Segmentation for Reference Pictures}
\begin{figure}[h]
	\centering
	\includegraphics[width=175mm]{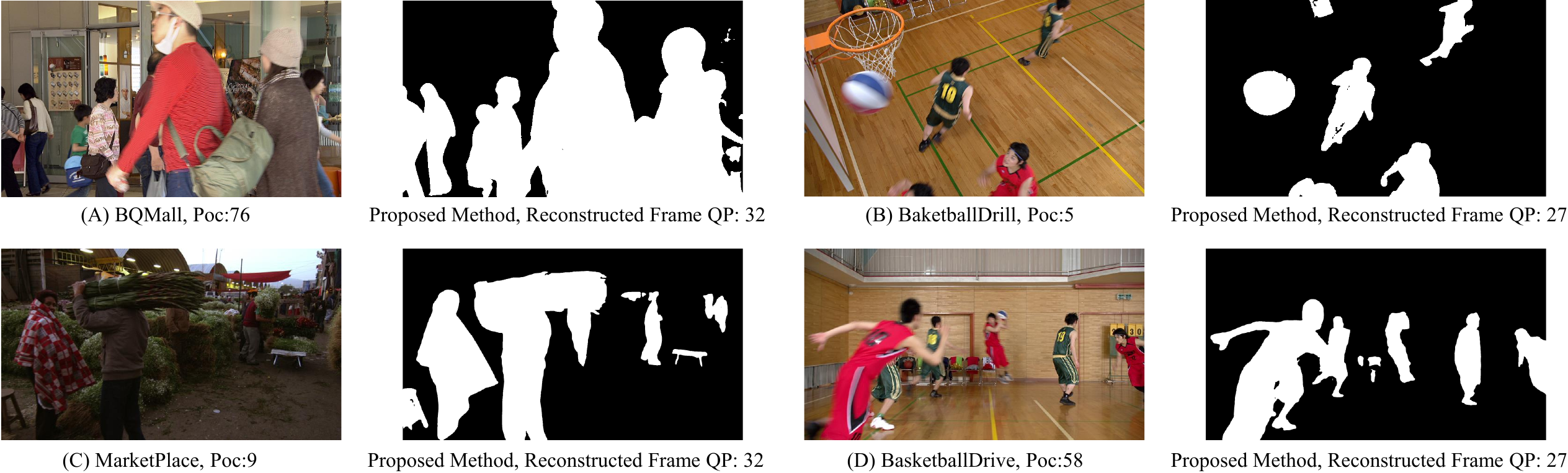}
	\caption{Visualization of segmentation masks of CTC test sequences segmented by our proposed segmentation strategy in the actual coding process.}	
	\label{fig:segmentation}
\end{figure}

Here we show some segmentation masks of CTC test sequences (Fig.~\ref{fig:segmentation}) segmented by our proposed segmentation strategy in the actual coding process.

\section{Limitation and Potential}
\subsection{Visualization of Failure Segmentation Cases}
\begin{figure}[h]
	\centering
	\includegraphics[width=180mm]{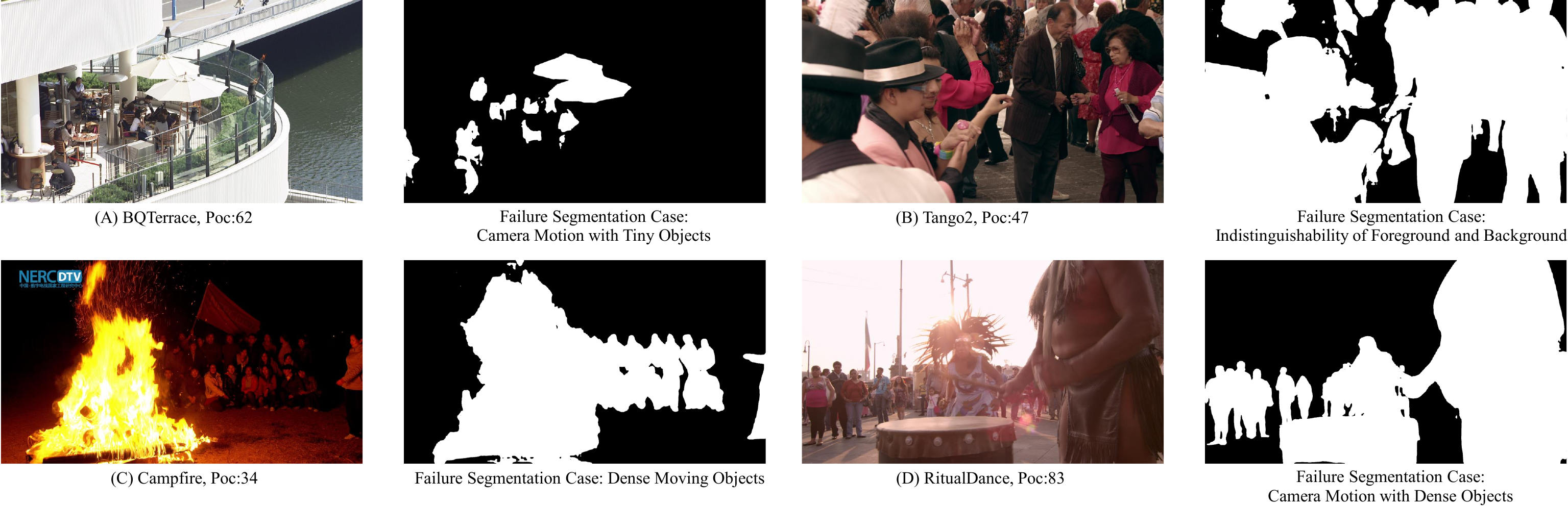}
	\caption{Failure segmentation cases on some sequences with special cases (extreme motion, serious occlusion, large resolution).}	
	\label{fig:Failurecase}
\end{figure}

Supplement Section  \uppercase\expandafter{\romannumeral9}.D 1), here we show some failure segmentation cases of CTC test sequences in Fig.~\ref{fig:Failurecase}.

\begin{figure}[h]   	
	\centering
	\small
	\includegraphics[width=170mm]{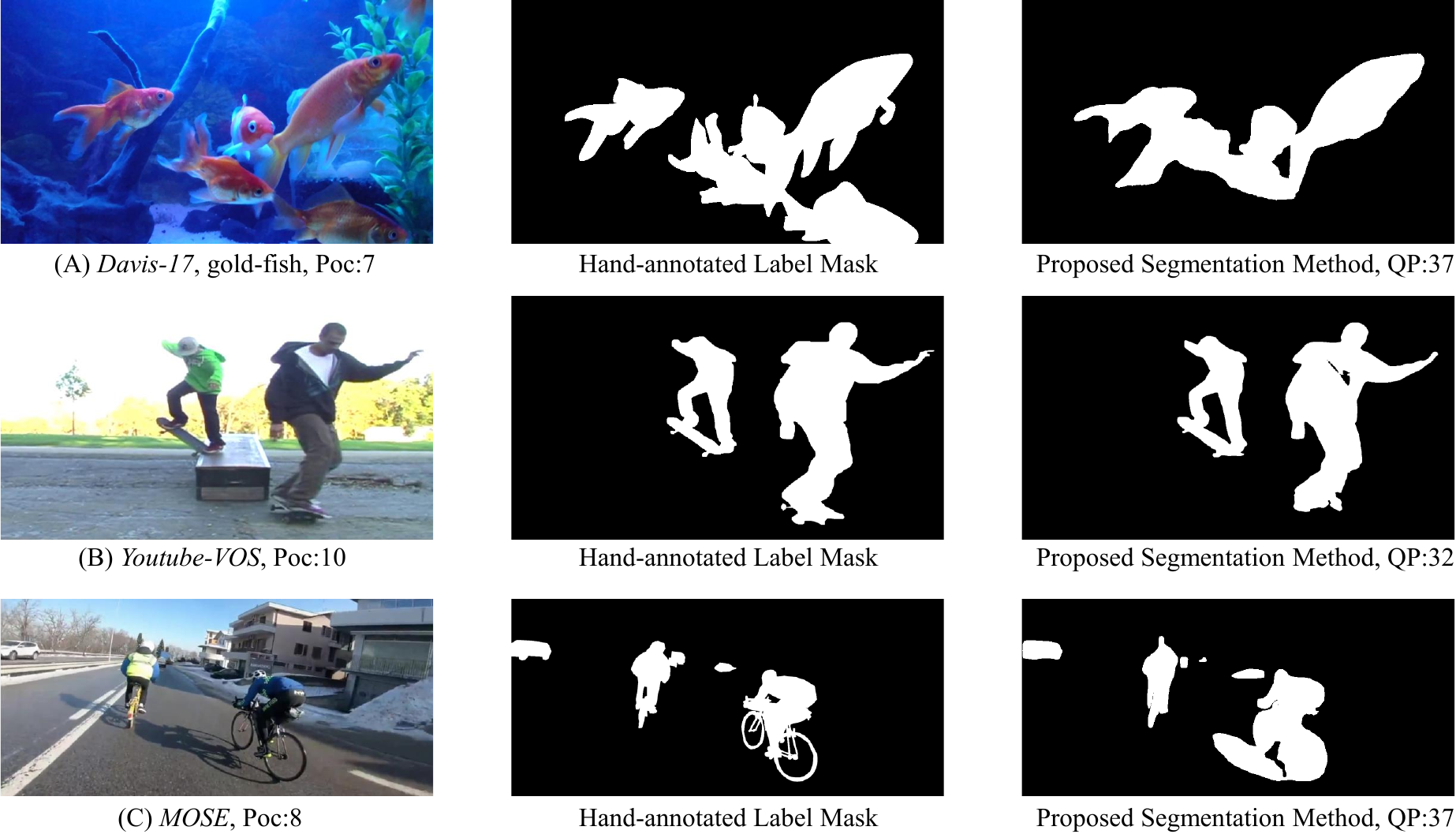}
	\caption{The difference of segmentation results of some selected test sequences of \textit{Davis-17}\cite{caelles20182018}, \textit{Youtube-VOS}\cite{{xu2018youtube}}, \textit{MOSE}\cite{ding2023mose}.}
	\label{fig:difference}
\end{figure}

\subsection{Setting of Potential Exploration Experiments}
Supplement Table \uppercase\expandafter{\romannumeral9} (BD-rate results of our proposed SAIP method compared to VTM-12.0 on the test sequences of VOS's train and validation dataset), here we detail the setting of the selected test sequences of VOS's train and validation dataset. For each benchmark (\textit{Davis-17}\cite{caelles20182018}, \textit{Youtube-VOS}\cite{{xu2018youtube}}, \textit{MOSE}\cite{ding2023mose}), we select 30 videos with
multiple moving objects. The selected sequences and the number of objects of each sequence (in parentheses) are shown as follows.

\textit{Davis-17}\cite{caelles20182018} (99 objects in total): bike-packing (2), boxing-fisheye (3), cat-girl (2), classic-car (3), color-run (3), crossing (3), dancing (3),	disc-jockey (3), dogs-jump (3),	dogs-scale (4), drone (4),	gold-fish (5), horsejump-high (2), horsejump-low (2), india (3), judo (2),	kid-football (2), lab-coat (3),	lindy-hop (8), loading (3),	longboard (5), motocross-bumps (2), motocross-jump (2),	night-race (2),	pigs (3), schoolgirls (7), sheep (5), shooting (3), tuk-tuk (3), train (4).	

\textit{Youtube-VOS}\cite{xu2018youtube} (120 objects in total): 0a7b27fde9 (2), 0a8c467cc3 (3), 0a23765d15 (4), 0ce06e0121 (2), 0043f083b5 (3), 00917dcfc4 (3), 011ac0a06f (5), 01c4cb5ffe (3), 01c76f0a82 (4), 0348a45bca (5), 0358b938c1 (4), 03c95b4dae (3), 07353b2a89 (4), 077d32af19 (4), 07c62c3d11 (3), 1122c1d16a (6), 1250423f7c (4), 165c42b41b (4), 19e061eb88 (4), 1aa3da3ee3 (4), 1af8d2395d (4), 1d3087d5e5 (5), 1ef8e17def (3), 1f7d31b5b2 (3), 1f8014b7fd (4), 21f4019116 (5), 351cfd6bc5 (5), 4122aba5f9 (5), c55784c766 (6), c4f5b1ddcc (6).

\textit{MOSE}\cite{ding2023mose} (295 objects in total): 78a1f4b1 (5),
55ae6921 (4), 072e7b3f (4), 23b9a3ea (14), a386eaa2 (6), 905d8740 (6), 93e22d48 (10), 25af2eeb (20), 6da17988 (8), fbfb6e30 (12), f6f8b4a6 (9), 662a1846 (9), 37af9ac1 (10), 8d0d641a (9), 831e0009 (8), 8c8b7913 (9), ba538072 (13), fabae187 (10), 82667d52 (10), ba0e4dd4 (8), 7be91184 (11), 14655f54 (10), adf443e1 (11), 2f477d05 (13), 2a2745e5 (11), e638e950 (13), 255f86ef (12), 86a2b59f (20), 23e046fb (10), 06ea172e (12).

Here we show the difference of segmentation results of some selected test sequences of \textit{Davis-17}\cite{caelles20182018}, \textit{Youtube-VOS}\cite{{xu2018youtube}}, \textit{MOSE}\cite{ding2023mose} in Fig.~\ref{fig:difference}\,, including the results segmented by our proposed segmentation method, hand-annotated label segmentation mask provided by the open-source annotation of train and validation dataset.

\section{Mode Selection Results}
\begin{figure}[h]   	
	\centering
	\small
	\includegraphics[width=180mm]{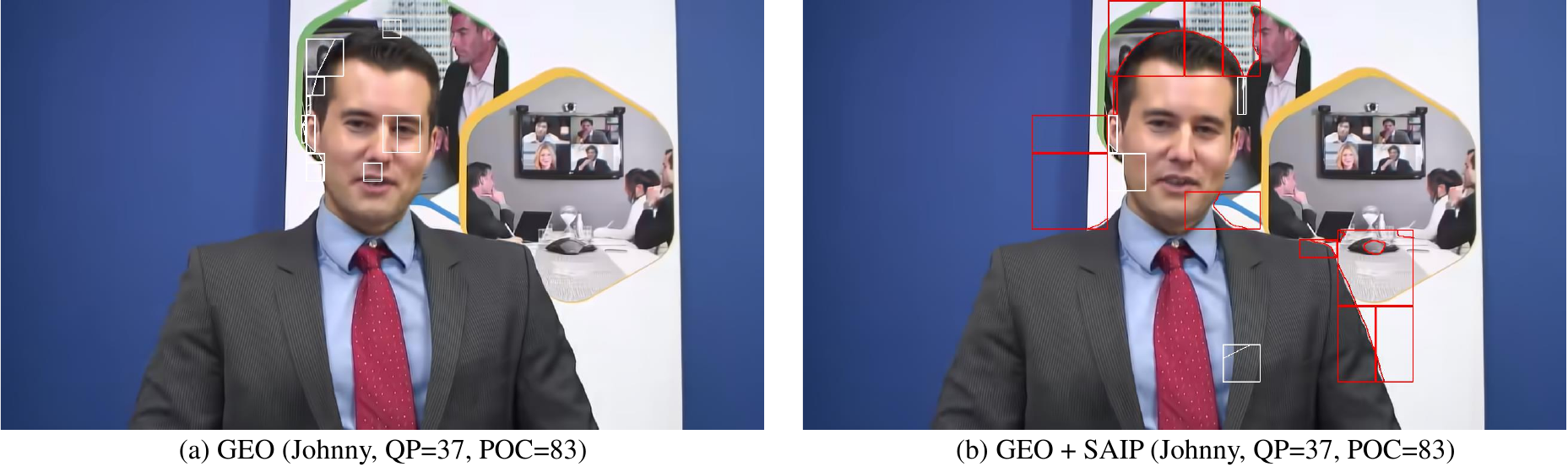}\vspace{1em}
	\includegraphics[width=180mm]{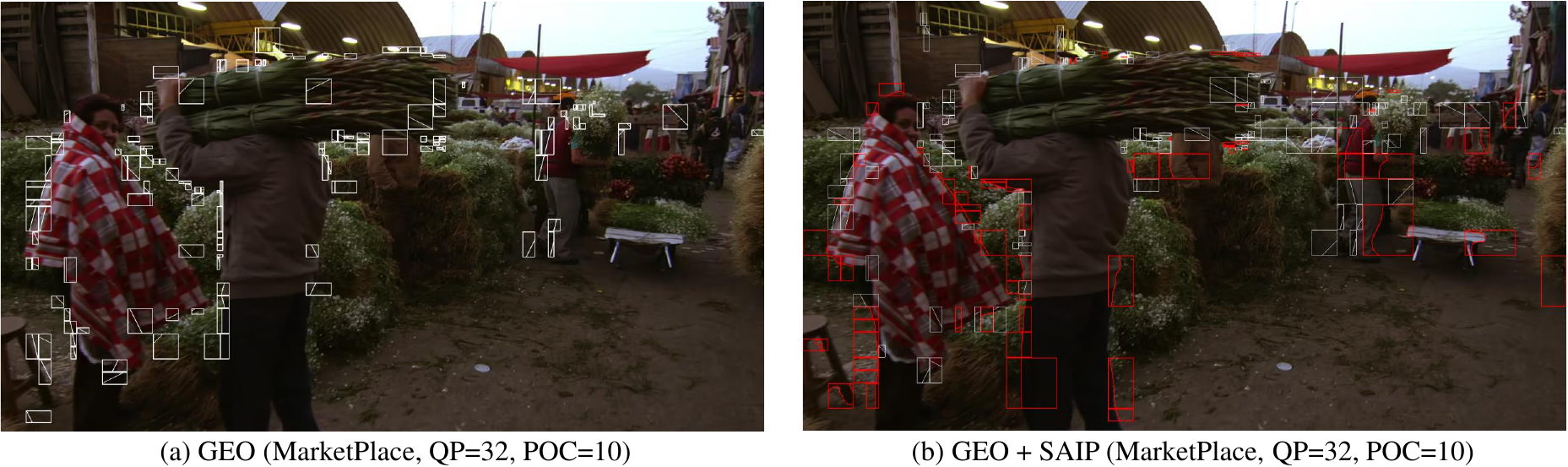}\vspace{1em}
	\includegraphics[width=180mm]{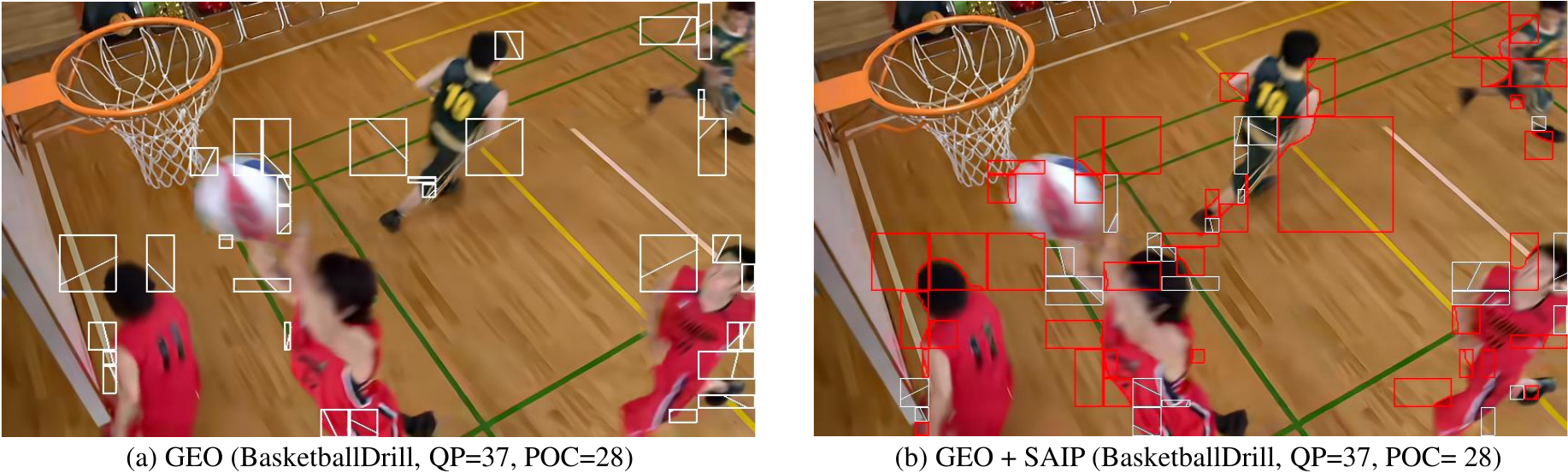}\vspace{1em}
	\includegraphics[width=180mm]{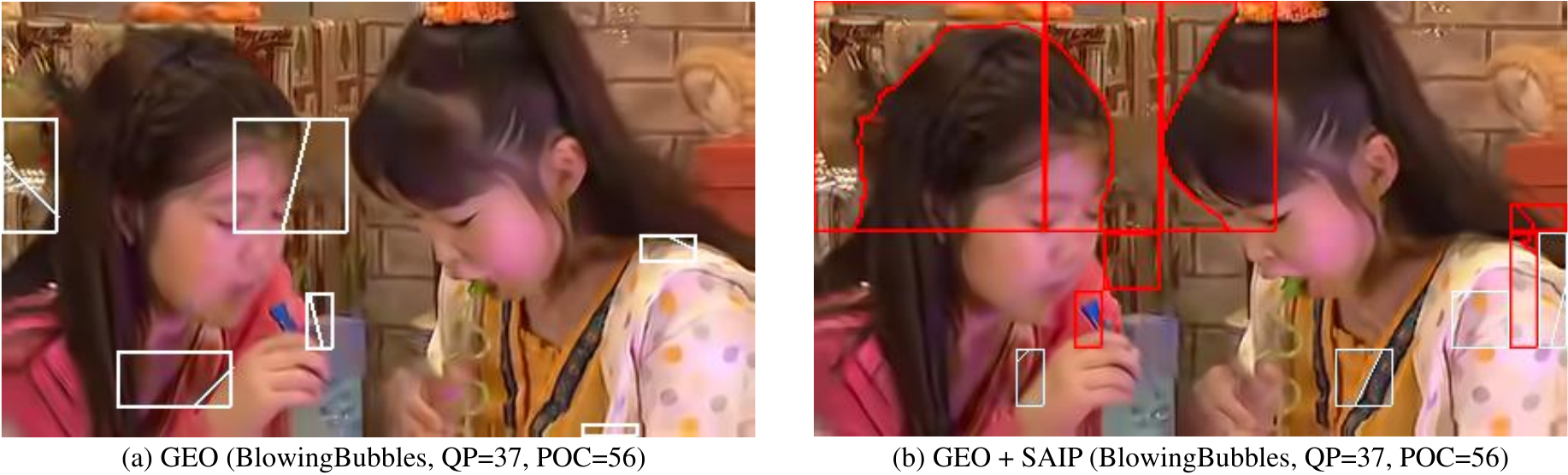}
	\caption{The selection results of different modes of CTC test sequences on VTM-12.0 (Configuration: LDB), where (a) uses the GEO mode (anchor: VTM-12.0), (b) uses the GEO mode and proposed SAIP mode. The white/red blocks indicate the block with GEO/SAIP selected.}
	\label{fig:srp1}
\end{figure}

In Fig. 9, we show the selection results of different modes (GEO, SAIP) on the BQMall sequence. Here we show more selection results on other sequences of CTC for the comparison of different modes (GEO, SAIP) in Fig.~\ref{fig:srp1}\,.

%

\bibliographystyle{IEEEtran}
\bibliography{IEEEexample}